\documentclass{aa}

\usepackage{txfonts}
\usepackage{graphicx}
\usepackage{amssymb}

\begin{document}

\title{Westerbork \ion{H}{i} observations of high-velocity clouds near M31 and M33} 

\author{T. Westmeier\inst{1,2} \and R. Braun\inst{2} \and D. Thilker\inst{3}}
\offprints{T. Westmeier,\\ \email{twestmei@astro.uni-bonn.de}}
\institute{Radioastronomisches Institut der Universit\"at Bonn, Auf dem
H\"ugel 71, 53121 Bonn, Germany \and ASTRON, P.O. Box 2, 7990 AA Dwingeloo,
Netherlands \and Center for 
Astrophysical Sciences, Johns Hopkins University, 3400 North Charles Street,
Baltimore, MD 21218, U.S.A.}

\date{Received 11 February 2005 / Accepted 28 February 2005}

\abstract{We have undertaken high-resolution follow-up of a sample of high
  velocity \ion{H}{i} clouds apparently associated with \object{M31}. Our sample
  was chosen from the population of high-velocity clouds (HVCs)
  detected out to 50~kpc projected radius of the Andromeda Galaxy by Thilker
  et al. (\cite{Thilker04}) with the Green Bank Telescope. Nine pointings were
  observed with the Westerbork Synthesis Radio Telescope to determine the
  physical parameters of these objects and to find clues to their origin. One
  additional pointing was directed at a similar object near \object{M33}.
  
  At $2'$ resolution we detect 16 individual HVCs around \object{M31} and 1 HVC 
  near \object{M33} with typical \ion{H}{i} masses of a few times $10^5 \, M_{\odot}$ 
  and sizes of the order of 1~kpc. 
  Estimates of the dynamical and virial masses of some of the HVCs indicate 
  that they are likely gravitationally dominated by additional 
  mass components such as dark matter or ionised gas. Twelve of the clouds are 
  concentrated in an area of only $1^{\circ} \times 1^{\circ}$ at a projected 
  separation of less than 15~kpc from the disk of \object{M31}. This HVC complex has a 
  rather complicated morphological and kinematical structure and partly overlaps 
  with the giant stellar stream of \object{M31}, suggesting a tidal origin. Another 
  detected feature is in close proximity, in both position and velocity, with 
  \object{NGC~205}, perhaps also indicative of tidal processes. Other HVCs in our survey 
  are isolated and might represent primordial, dark-matter dominated clouds.

  } 

\maketitle

\section{\label{sect_introduction}Introduction}

\begin{figure}
  \begin{center}
    \includegraphics[width=\linewidth]{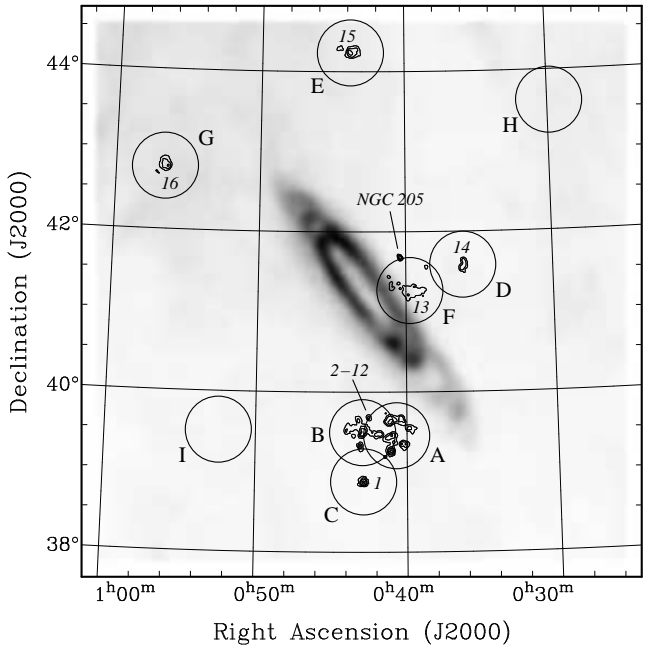}
  \end{center}
  \caption{Overview of the 9 fields (black circles, labelled with capital
  letters) observed around \object{M31} with the WSRT. The grayscale image displays the
  \ion{H}{i} emission of \object{M31} observed by Thilker et al. (2004) with the
  GBT. The contours show our \ion{H}{i} data of the 16 high-velocity clouds
  detected with the WSRT (labelled with numbers) and of \object{NGC~205}. In fields H
  and I no reliable signal of high-velocity \ion{H}{i} emission was
  found at the resolution and sensitivity of our WSRT data. \label{fig_field}}
\end{figure}

High-velocity clouds (HVCs) were discovered by Muller et al. (\cite{Muller63})
in the 21-cm line emission of neutral atomic hydrogen. They are identified by
their peculiar radial velocities which are incompatible with participation
in the normal rotation of our Galaxy. HVCs can be found all across the sky in
a variety of different sizes and shapes (Wakker \& van Woerden
\cite{Wakker97}). Some HVCs form large and coherent complexes spanning tens of
degrees in size. Others are comparatively small and isolated. The origin of
HVCs is still under debate. In the decades since their discovery many
different mechanism have been proposed to explain the observed population of
HVCs (see Wakker \& van Woerden \cite{Wakker97} for a detailed review). One of
the most prominent hypotheses puts forward the Galactic fountain (Shapiro \& Field
\cite{Shapiro76}) as a means of forming HVCs. In this scenario, hot gas ejected
by the Galactic fountain would cool down and recombine high above the Galactic
plane. The condensed gas clouds, falling back ballistically towards the
Galactic plane, would then appear as neutral high-velocity clouds. Other
scenarios suggest that the gas of HVCs was torn out of dwarf galaxies during
close encounters with the Milky Way by tidal forces or by ram-pressure
interaction. At least two prominent structures classified as HVCs, the
Magellanic Stream (Mathewson et al. \cite{Mathewson74}) and the Leading Arm
(Gardiner \& Noguchi \cite{Gardiner96}; Putman et al. \cite{Putman98}), were
formed in this way.

Another hypothesis that has been discussed extensively in recent years postulates 
that HVCs are primordial gas clouds left over from the formation of the galaxies
in the Local Group. This scenario was first suggested by Oort (\cite{Oort66})
and recently revived in the framework of cold dark matter (CDM) cosmology by
Blitz et al. (\cite{Blitz99}), who found the kinematic properties of most 
HVCs to be consistent with a distribution throughout the entire Local
Group. They proposed that HVCs might represent the gaseous counterparts of the
so-called ``missing'' dark matter satellites which were predicted by CDM
structure formation scenarios (Klypin et al. \cite{Klypin99}; Moore et
al. \cite{Moore99}; Kravtsov et al. \cite{Kravtsov04}). On the basis of the
Leiden/Dwingeloo Survey (LDS) of Galactic neutral hydrogen (Hartmann \& Burton
\cite{Hartmann97}) Braun \& Burton (\cite{Braun99}) identified a sub-sample of
66 compact high-velocity clouds (CHVCs) with angular sizes of less than
$2^{\circ}$. The kinematic parameters of the CHVCs suggested an unvirialized
population within the Local Group potential with distances that might extend
to the turn-around radius of about $1 \; {\rm Mpc}$, representing the
primordial dark-matter mini-haloes which have not yet been accreted by one of
the large galaxies.

Recent observations have raised doubt on an extended Local Group scenario. If
the observed CHVC population had a typical distance as large as 1~Mpc,
implying a typical \ion{H}{i} mass of $10^8 \, M_{\odot}$, then we would
expect to detect a similar population of \ion{H}{i} clouds in other galaxy
groups. However, all attempts to detect such an intra-group population in
nearby galaxy groups, employing \ion{H}{i} mass limits of about $10^7 \,
M_{\odot}$, have failed (Zwaan \cite{Zwaan01}; Braun \& Burton \cite{Braun01};
Pisano et al. \cite{Pisano04}). In addition, H$\alpha$ observations towards a
number of CHVCs (Tufte et al. \cite{Tufte02}; Putman et al. \cite{Putman03})
have been interpreted to imply distances of the order of only $10 \; {\rm
kpc}$ from the Galaxy, where the Galactic radiation field is strong enough to
account for the detected H$\alpha$ intensities. All these observations suggest
that the CHVCs are distributed throughout the circum-Galactic environment, more 
closely associated with the Milky Way rather than being concentrated near the 
Local Group turn-around radius.

\begin{figure}
  \begin{center}
    \includegraphics[width=\linewidth]{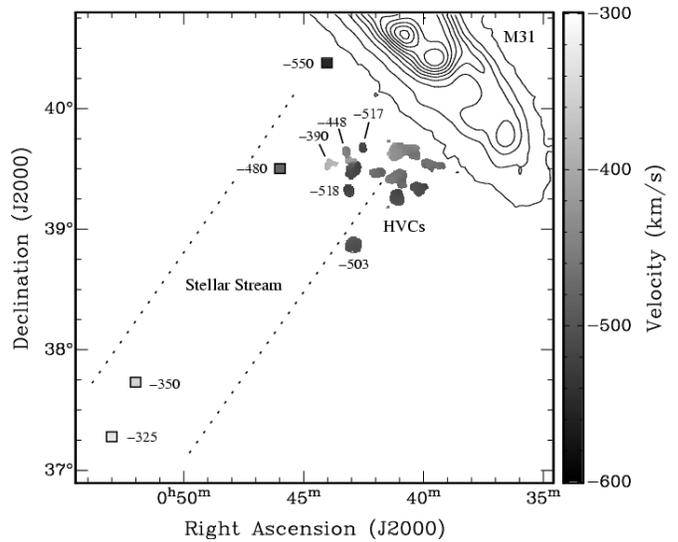}
  \end{center}
\caption{The map compares the position and heliocentric velocitiy of the giant
stellar stream of \object{M31} with the nearby high-velocity clouds. The squares
indicate the positions of the radial velocity measurements along the stream
(outlined by the dotted lines) by Ferguson et al. (\cite{Ferguson04}). For
comparison we have inserted the first moment maps of the HVCs from our data.
The GBT data of the \ion{H}{i} disk of \object{M31} from Thilker et
al. (\cite{Thilker04}) is shown for orientation in black contours in the upper
right corner of the map.  The numbers denote the heliocentric radial
velocities of individual features.
\label{fig_stellarstream}}
\end{figure}

The closest large spiral galaxy comparable to the Milky Way is the Andromeda
galaxy (\object{M31}), making it an ideal target to verify the circum-galactic nature
of high-velocity clouds observationally. Thilker et al. (\cite{Thilker04})
used the Green Bank telescope (GBT) to observe a field of $7^{\circ} \times
7^{\circ}$ around \object{M31} in the 21-cm line of neutral hydrogen. They discovered
a population of \ion{H}{i} clouds (with \ion{H}{i} masses between $10^5$ and $10^6 
\, M_{\odot}$, in the case of Davies' Cloud about $10^7 \, M_{\odot}$) which 
are likely to be the counterparts of the high-velocity clouds around our Milky 
Way. A similar GBT survey of a $5^{\circ} \times 5^{\circ}$ field around \object{M33} 
by Thilker et al. (in prep.)  covered another anomalous-velocity \ion{H}{i} 
cloud close to the disk of \object{M33}. This cloud was already described by Huchtmeier 
(\cite{Huchtmeier78}) and has properties similar to the HVCs found in the 
direction of \object{M31}. Many of the clouds in the direction of \object{M31} and \object{M33} are rather 
compact and barely resolved with the GBT. Moreover, complex line profiles in 
some directions indicate the presence of compact substructure which is unresolved 
in the GBT data. Therefore, we re-observed some of the HVCs near \object{M31} and \object{M33} in 
\ion{H}{i}, using the Westerbork Synthesis Radio Telescope (WSRT). The aim of our 
interferometric observations was to study the HVCs and their compact substructure 
with higher angular resolution and to determine their physical parameters.

\begin{table*}
  \caption{Physical parameters of the high-velocity clouds found near \object{M31} and 
  \object{M33}. The columns give the cloud name, the right
  ascension $\alpha$ and declination $\delta$, the peak column density
  $N_{\rm HI}$ at $2'$ resolution, the heliocentric radial velocity $v_{\rm hel}$, the average
  line width (FWHM) of the \ion{H}{i} lines $\Delta v$, the cloud diameter $D$
  (FWHM), the \ion{H}{i} mass $M_{\rm HI}$, the \ion{H}{i} mass in the GBT data 
  of Thilker et al. (\cite{Thilker04}) $M_{\rm HI}^{\mathrm{GBT}}$, the dynamical 
  mass $M_{\rm dyn}$, the virial mass $M_{\rm vir}$, the average \ion{H}{i} volume 
  density $n_{\rm HI}$, the value of the velocity gradient $\delta v$, the position angle 
  of the velocity gradient $\varphi$ (measured in the mathematically negative 
  sense with $0^{\circ}$ being north), and the ellipticity $e$. Distances of 
  780~kpc for \object{M31} and 800~kpc for \object{M33} were assumed for the clouds to calculate 
  their sizes, masses, and densities. In the last two rows we have calculated 
  the mean value $\langle x \rangle$ and the standard deviation $\sigma_x$ of 
  some parameters of M31~HVCs~1--16. \label{tab_properties}}
\begin{center}
\begin{tabular}{lrrrrrrrrrrrrrr}
\hline \hline
cloud name & $\alpha$ & $\delta$ & $N_{\mathrm{HI}}$ & $v_{\mathrm{hel}}$ & $\Delta v$ & $D$ & $M_{\mathrm{HI}}$ & $M_{\mathrm{HI}}^{\mathrm{GBT}}$ & $M_{\mathrm{dyn}}$ & $M_{\mathrm{vir}}$ & $n_{\mathrm{HI}}$ & $\delta v$ & $\varphi$ & $e$ \\
       &  (J2000)  &  (J2000)  &  ($10^{19}$  &  \multicolumn{2}{c}{($\mathrm{km \, s}^{-1}$)}  &  (kpc)  &  \multicolumn{2}{c}{($10^5 \, M_{\odot}$)}  &  \multicolumn{2}{c}{($10^5 \, M_{\odot}$)}  &  ($10^{-2}$ & ($\mathrm{km \, s}^{-1}$ & ($^{\circ}$) &  \\
 & & & $\mathrm{cm}^{-2}$) & & & & & & & & $\mathrm{cm}^{-3}$) & $\mathrm{kpc}^{-1}$) & & \\
\hline
\object{M31 HVC 1}  & $0^{\rm h}42^{\rm m}53^{\rm s}$ & $38^{\circ}52'$ & $11.1$    & $-503$ & $17.3$ & $0.72$ & $5.2$ & $4.8$ & $>\!86$ & $230$     & $10.9$	 & $ 7.7$ & $349$ & $0.24$     \\
\object{M31 HVC 2}  & $0^{\rm h}41^{\rm m}03^{\rm s}$ & $39^{\circ}16'$ & $ 9.5$    & $-511$ & $29.6$ & $0.75$ & $5.0$ &       & $>\!44$ & $690$     & $ 9.3$	 & $ 9.7$ & $188$ & $0.12$     \\
\object{M31 HVC 3}  & $0^{\rm h}40^{\rm m}11^{\rm s}$ & $39^{\circ}21'$ & $ 3.4$    & $-509$ & $18.5$ & $0.93$ & $2.2$ &       &	 & $330$     & $ 2.1$	 & $ 8.4$ & $239$ & $0.28$     \\
\object{M31 HVC 4}  & $0^{\rm h}43^{\rm m}07^{\rm s}$ & $39^{\circ}20'$ & $ 4.2$    & $-518$ & $21.0$ & $0.56$ & $1.3$ &       &	 & $260$     & $ 5.8$	 & $ 4.1$ & $ 98$ & $0.22$     \\
\object{M31 HVC 5}  & $0^{\rm h}42^{\rm m}51^{\rm s}$ & $39^{\circ}31'$ & $ 5.8$    & $-512$ & $25.3$ & $1.13$ & $6.0$ &       &	 &	     & $ 3.3$	 &	  &	  & $0.48$     \\
\object{M31 HVC 6}  & $0^{\rm h}42^{\rm m}31^{\rm s}$ & $39^{\circ}41'$ & $ 1.8$    & $-517$ & $17.3$ & $0.53$ & $0.6$ &       & $>\!19$ & $170$     & $ 3.2$	 & $ 4.1$ & $334$ & $0.42$     \\
\object{M31 HVC 7}  & $0^{\rm h}41^{\rm m}03^{\rm s}$ & $39^{\circ}26'$ & $ 3.7$    & $-464$ & $20.3$ & $1.27$ & $3.5$ &       &	 &	     & $ 1.3$	 & $10.7$ & $280$ & $0.40$     \\
\object{M31 HVC 8}  & $0^{\rm h}41^{\rm m}47^{\rm s}$ & $39^{\circ}28'$ & $ 3.0$    & $-476$ & $32.1$ & $0.92$ & $2.1$ &      & $>\!730$ &	     & $ 2.1$	 & $15.5$ & $291$ & $0.36$     \\
\object{M31 HVC 9}  & $0^{\rm h}43^{\rm m}13^{\rm s}$ & $39^{\circ}39'$ & $ 1.8$    & $-448$ & $19.1$ & $0.69$ & $0.8$ &       &	 & $260$     & $ 1.9$	 & $12.7$ & $252$ & $0.51$     \\
\object{M31 HVC 10} & $0^{\rm h}41^{\rm m}03^{\rm s}$ & $39^{\circ}38'$ & $ 3.9$    & $-432$ & $30.2$ & $1.20$ & $5.0$ &      & $>\!810$ &           & $ 2.3$    & $14.7$ & $262$ & $0.38$     \\
\object{M31 HVC 11} & $0^{\rm h}43^{\rm m}32^{\rm s}$ & $39^{\circ}27'$ & $ 0.9$    & $-454$ & $30.2$ & $0.43$ & $0.2$ &       &         & $410$     & $ 2.0$    &        &       & $0.26$     \\
\object{M31 HVC 12} & $0^{\rm h}43^{\rm m}49^{\rm s}$ & $39^{\circ}33'$ & $ 1.3$    & $-390$ & $30.8$ & $1.5 $ & $1.5$ &       &         &           & $ 0.3$    &        &       &            \\
\object{M31 HVC 13} & $0^{\rm h}39^{\rm m}35^{\rm s}$ & $41^{\circ}15'$ & $ 1.3$    & $-211$ & $22.8$ & $2.6 $ & $3.5$ & $8.7$ &         &           & $ 0.2$    & $ 4.6$ & $233$ & $0.65$     \\
\object{M31 HVC 14} & $0^{\rm h}36^{\rm m}04^{\rm s}$ & $41^{\circ}34'$ & $ 2.9$    & $-512$ & $24.0$ & $1.11$ & $2.5$ & $3.0$ &         &           & $ 1.4$    &        &       & $0.61$     \\
\object{M31 HVC 15} & $0^{\rm h}43^{\rm m}50^{\rm s}$ & $44^{\circ}14'$ & $ 4.8$    & $-273$ & $26.5$ & $1.07$ & $5.2$ & $5.2$ &         & $790$     & $ 3.3$    & $ 4.5$ & $176$ & $0.37$     \\
\object{M31 HVC 16} & $0^{\rm h}56^{\rm m}25^{\rm s}$ & $42^{\circ}47'$ & $ 3.4$    & $-164$ & $25.9$ & $1.26$ & $3.9$ & $4.4$ &         & $890$     & $ 1.5$    & $ 8.9$ & $115$ & $0.43$     \\
\hline
\object{M31 HVC A}  & $0^{\rm h}40^{\rm m}23^{\rm s}$ & $39^{\circ}41'$ & $ 2.5$    & $-455$ & $32.1$ & $1.04$ & $1.9$ &       &         &           & $ 1.3$    &        &       & $0.56$     \\
\object{M31 HVC B}  & $0^{\rm h}39^{\rm m}54^{\rm s}$ & $39^{\circ}33'$ & $ 1.9$    & $-468$ & $39.5$ & $0.91$ & $1.3$ &       &         &           & $ 1.3$    &        &       & $0.57$     \\
\hline
\object{M33 HVC 1}  & $1^{\rm h}32^{\rm m}33^{\rm s}$ & $29^{\circ}34'$ & $ 5.2$    & $-147$ & $17.3$ & $1.33$ & $7.8$ &       &         &           & $ 2.6$    &        &       & $0.43$     \\
\hline
$\langle x \rangle$  &                       &                 & $ 3.9$    &        & $24.4$ & $1.04$ & $3.0$ &       &         & $450$     & $ 3.2$    & $ 8.8$ &       & $0.38$     \\
$\sigma_x$           &                       &                 & $\pm 2.8$ &     & $\pm 5.1$ & $\pm 0.52$ & $\pm 1.9$ & &       & $\pm 270$ & $\pm 3.0$ & $\pm 4.1$ &    & $\pm 0.15$ \\
\hline
\end{tabular}
\end{center}
\end{table*}

In this paper we present the results of these WSRT observations and discuss
the various possible origins of the detected HVCs in the neighbourhood of \object{M31}
and \object{M33}. After describing the sample selection in
Sect.~\ref{sect_sampleselection}, we specify the technical parameters of the
observations and the data reduction procedure in
Sect.~\ref{sect_observations}. In Sect.~\ref{sect_results} the results of our
observations are presented. A discussion of our results and their implications
can be found in Sect.~\ref{sect_discussion}. Finally, Sect.~\ref{sect_summary}
gives a summary of our results and conclusions.

\section{\label{sect_sampleselection}Sample selection}

Based on the GBT observations of Thilker et al. (\cite{Thilker04}) and Thilker
et al. (in prep.) we selected 9 fields around \object{M31} and 1 field near \object{M33} for
follow-up observations of HVCs with the WSRT. The 9 fields around \object{M31} were
centred on the brightest and most conspicuous clouds and cover about half of
the population of HVCs discovered by Thilker et al. (\cite{Thilker04}). The
targets cover the different morphological types found in the GBT data and
include HVCs with both large and small angular separations from the disk of
\object{M31}. The most prominent cloud near \object{M31}, \object{Davies' Cloud}, was already studied in
great detail (Davies \cite{Davies75}; de Heij et al. \cite{deHeij02}) so 
we did not include it in our survey. An overview of the 9 fields around \object{M31}
can be found in Fig.~\ref{fig_field}.

The field near \object{M33} is located about $1^{\circ}$ south of the centre of the
galaxy. It is centred on a faint \ion{H}{i} cloud outside the disk of \object{M33}
which is prominent in the GBT data. The cloud is not isolated but appears to be
connected with the \ion{H}{i} disk of \object{M33} by a faint gas bridge.

\begin{figure*}
  \begin{center}
    \includegraphics[width=\linewidth]{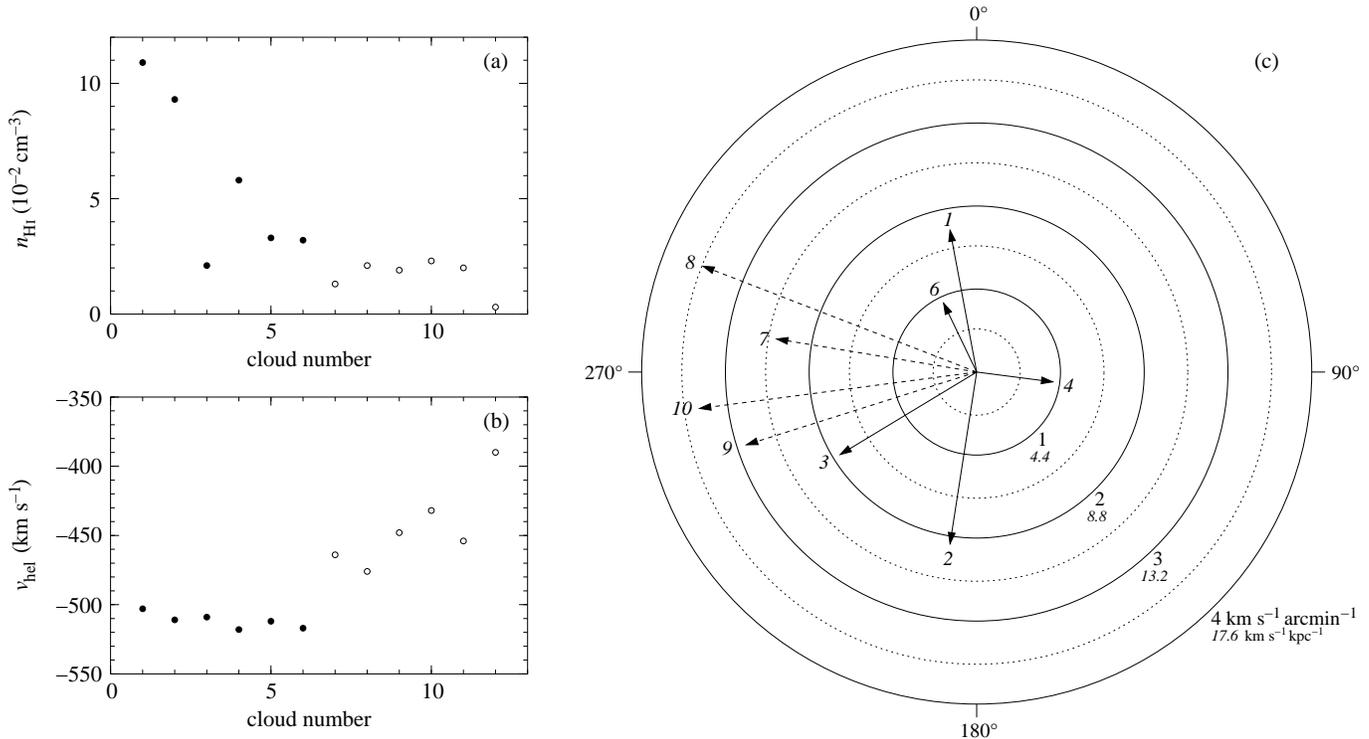}
  \end{center}
\caption{\textbf{(a)} \ion{H}{i} volume densities $n_{\mathrm{HI}}$ and
\textbf{(b)} heliocentric radial velocities $v_{\mathrm{hel}}$ of
M31~HVCs~1--12.  M31~HVCs~1--6 ($\bullet$) have similar radial velocities, and
they are much more condensed with typically higher densities. M31~HVCs~7--12
($\circ$), in contrast, are more diffuse with lower densities and rather 
different radial velocities.  In \textbf{(c)} we have plotted the value of the
velocity gradient across some of the clouds (labelled with italic numbers) in
$\mathrm{km \, s}^{-1} \, \mathrm{arcmin}^{-1}$ (labelled from 1 to 4) against
the direction of the gradient in degrees. It turns out that the gradients
across M31~HVCs~1--6 (solid arrows) have arbitrary orientations whereas the
typically larger gradients across M31~HVCs~7--10 (dashed arrows) are all
oriented in about the same direction. Note that for M31~HVCs~5, 11, and 12 we 
could not obtain a meaningful fit for the velocity gradient.}
  \label{fig_complex}
\end{figure*}

\section{\label{sect_observations}Observations \& data reduction}

The 10 selected fields were observed in the 21-cm line emission of \ion{H}{i},
using the WSRT in the Maxi-Short configuration. This configuration provides
particularly good sampling of the shortest East-West baselines (36, 54, 72 and
90~m). A complete 12-hour track for each field was preceded and followed by
observations of the external calibrator sources CTD93 and 3C147. The
correlator provided 1024 spectral channels for each of the two linear
polarisations. To improve the signal-to-noise ratio, the 1024 channels were
equally divided among two independent IF bands tuned to the same central
frequency, resulting in 512 channels for the final spectrum. By later
averaging these two IF bands we are able to decrease the digitization noise by
a factor $\sqrt{2}$. As a result, the noise level in the final images is about
5\% lower in comparison with the normal single-IF mode. With a bandwidth of
5~MHz the corresponding velocity resolution is about $2 \; \mathrm{km \,
s}^{-1}$.

The data reduction was done with the AIPS software package. After flagging the
data affected by radio frequency interference or by shadowing of the
telescopes we carried out the standard bandpass, gain, and flux calibration,
using the two external calibrators. To further improve the gain calibration,
we self-calibrated on the continuum sources in each field in an iterative
procedure. This provided us with a clean component model of the continuum
emission which was then subtracted from the visibility data.

As the emission of most HVCs in our survey was quite faint and extended, we
applied a Gaussian $uv$ taper, declining to 30\% amplitude at 1.25~k$\lambda$
radius, before producing the final, cleaned images. The resulting synthesised
beam has a FWHM of about $2'$, corresponding to a spatial resolution of about 
450~pc at the distance of \object{M31}. For the brightest and most compact clouds we
also produced images with a higher angular resolution of $30''$ FWHM (using a
$uv$ taper radius of 5~k$\lambda$). The velocity resolution was also smoothed
to $12 \; \mathrm{km \, s}^{-1}$ in all spectra to further increase the
signal-to-noise ratio, given that the typical detected linewidth was about $25
\; \mathrm{km \, s}^{-1}$. For the $2'$ resolution maps, at $12 \; \mathrm{km
\, s}^{-1}$, we obtain an {\sc RMS} sensitivity of about 1~mJy/beam towards
the centre of the field, corresponding to a brightness temperature {\sc RMS}
of 40~mK. This implies a $1 \sigma$ \ion{H}{i} column density sensitivity of
$9 \cdot 10^{17} \; \mathrm{cm}^{-2}$ for emission filling the $2'$ beam.

\begin{figure*}
  \begin{center}
    \includegraphics[width=\linewidth]{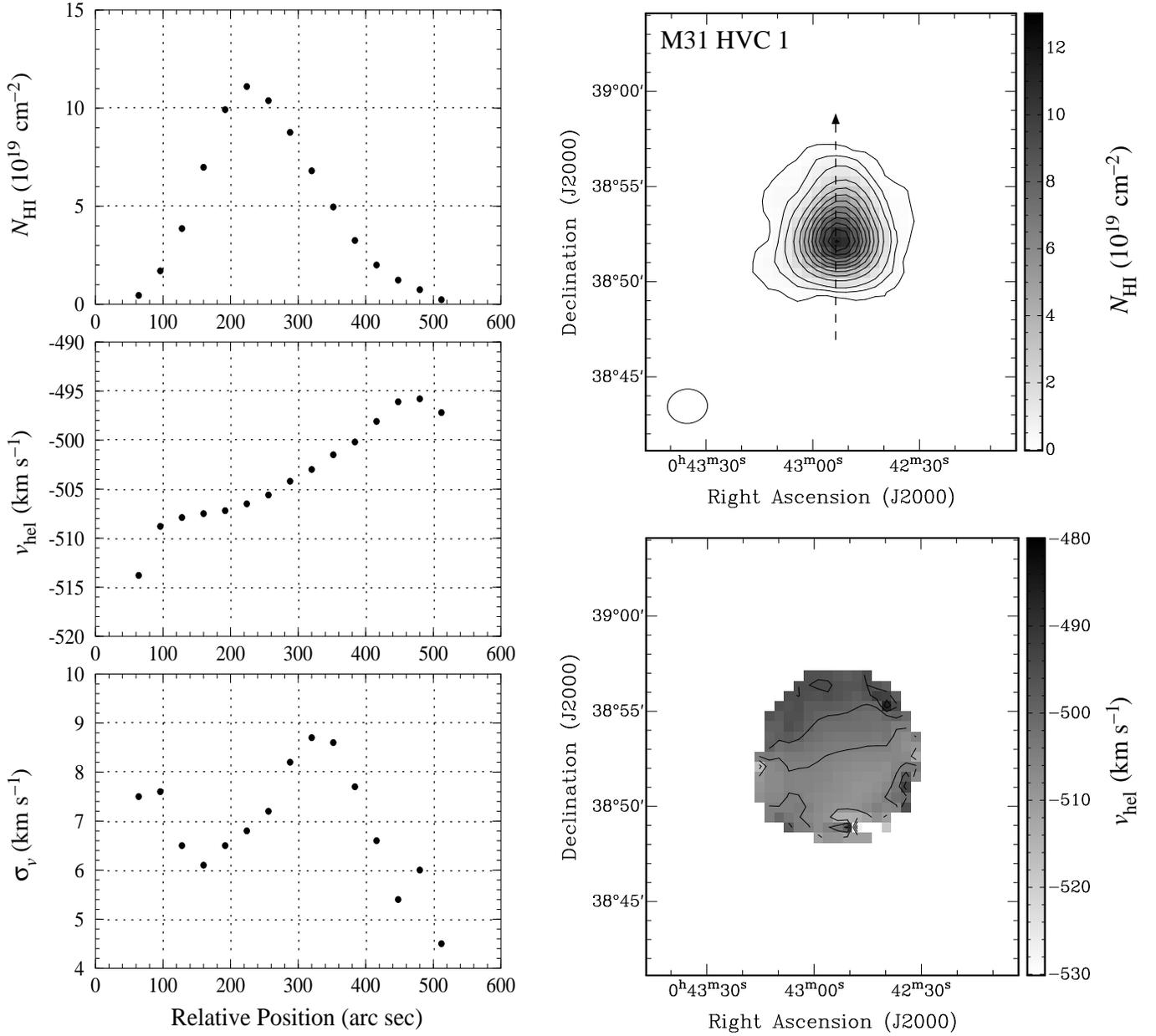}
  \end{center}
\caption{\object{M31~HVC~1}. The upper right map displays the \ion{H}{i} column
density. The contours are drawn at $1 \cdot 10^{18} \; \mathrm{cm}^{-2}$, $5
\cdot 10^{18} \; \mathrm{cm}^{-2}$, and from $1 \cdot 10^{19} \;
\mathrm{cm}^{-2}$ in steps of $1 \cdot 10^{19} \; \mathrm{cm}^{-2}$. The lower
right map shows the heliocentric radial velocity. The contours are separated
by $5 \; \mathrm{km \, s}^{-1}$. The diagrams on the left hand side show from
top to bottom: the \ion{H}{i} column density, the heliocentric radial
velocity, and the velocity dispersion of the gas along the cut marked by the
arrow in the column density map.}
  \label{fig_ov1}
\end{figure*}

\section{\label{sect_results}Results}

In two of the nine fields around \object{M31} no high-velocity gas was detected. In
both cases the emission of the high-velocity gas is probably too faint and/or
diffuse to be detected in our WSRT data given the finite column density
sensitivity noted above ($9 \cdot 10^{17} \; \mathrm{cm}^{-2}$ over $12 \;
\mathrm{km \,s}^{-1}$). In all other fields we clearly detect the HVCs at more
than $5 \sigma$ significance. The individual clouds have been named
M$xy$~HVC~$z$, where $xy$ is the Messier number of the host galaxy and $z$ is
the catalogue number assigned to each HVC.

The observational parameters of all clouds are summarised in
Table~\ref{tab_properties}. Under the assumption of a constant distance of
780~kpc (Stanek \& Garnavich \cite{Stanek98}) we obtain a mean linear size of
the 16 HVCs around \object{M31} of about 1~kpc FWHM and a mean \ion{H}{i} mass of $3
\cdot 10^5 \, M_{\odot}$. For the HVC near \object{M33} we assumed a distance of
800~kpc (Lee et al. \cite{Lee02}) to determine the physical parameters.

The \ion{H}{i} column densities were derived from the zeroth moment of the
spectra under the assumption that the optical depth of the gas is
negligible. From the integrated column density
$N_{\mathrm{HI}}^{\mathrm{tot}}$ we can directly calculate the \ion{H}{i} mass
of each cloud via $M_{\mathrm{HI}} = m_{\mathrm{H}} \, d^2 \tan^2 \! \varphi
\, N_{\mathrm{HI}}^{\mathrm{tot}}$, where $m_{\mathrm{H}}$ is the mass of a
hydrogen atom, $d$ is the distance of the cloud, and $\varphi$ denotes the
angular size of a resolution element of the map. The observed \ion{H}{i}
masses are in the range of a few times $10^4 \, M_{\odot}$ to $6 \cdot 10^5 \,
M_{\odot}$. For the isolated clouds in the direction of \object{M31} we have also 
listed in Table~\ref{tab_properties} the \ion{H}{i} masses derived from the 
GBT data of Thilker et al. (\cite{Thilker04}). These values are in good 
agreement with our WSRT observations. Only for the most extended and diffuse 
cloud, \object{M31~HVC~13}, we detect only about 40\% of the total flux.

The diameters of the HVCs were determined by fitting a Gaussian to
the radial column density distribution. By dividing the \ion{H}{i} mass by the
volume defined by the FWHM of this Gaussian, we can estimate a mean \ion{H}{i}
volume density $n_{\mathrm{HI}}$ for each cloud under the additional
assumption of spherical symmetry. The derived densities are typically of the
order of a few times $10^{-2} \; \mathrm{cm}^{-3}$. These volume densities are 
between two and three orders of magnitude higher than the densities of 
$10^{-5} \ldots 10^{-4} \; \mathrm{cm}^{-3}$ expected for an extended galactic 
corona in which the HVCs might be embedded (Sembach et al. \cite{Sembach03}; 
Rasmussen et al. \cite{Rasmussen03}).

The mean radial velocities of the HVCs were calculated from the first moment
of the spectra. Most values are more negative than the \object{M31} systemic velocity
of $v_{\mathrm{hel}} \approx -300 \; \mathrm{km \, s}^{-1}$. However, this is
a selection effect, because most of the HVCs detected around \object{M31} belong to a
filamentary complex of clouds with similar radial velocities near the southern
edge of the galaxy. The HVC near \object{M33} has a slightly less negative radial
velocity with respect to the \object{M33} systemic velocity of $v_{\mathrm{hel}}
\approx -180 \; \mathrm{km \, s}^{-1}$. By fitting a Gaussian to the spectral
lines we extracted the line widths with typical values in the range of about
$20 \ldots 30 \; \mathrm{km \, s}^{-1}$ FWHM at $2'$ resolution. These values 
correspond to an upper limit for the kinetic temperature of the gas of the order 
of $10^4 \; \mathrm{K}$. Similar line widths are observed for the Galactic HVCs 
(e.g. Wakker \& van Woerden \cite{Wakker97}, Braun \& Burton \cite{Braun99}, 
Westmeier et al. \cite{Westmeier05b}), and they are typical for the 
warm neutral medium (Braun \cite{Braun97}). Thilker et al. (\cite{Thilker04}) 
already noted that the large line widths of the HVCs discovered in the GBT
survey of \object{M31} require a large fraction of undetected mass to allow for 
gravitational stability of the clouds. We address this problem in more detail
in the discussion in Sect.~\ref{sect_stability}.

We also determined the radial velocity gradients across some of the HVCs by
fitting a plane $P(x,y) = cx + dy +f$ to the first moment maps
$v_{\mathrm{hel}}(x,y)$. The chi-square minimisation provided us with values
for the parameters $c$, $d$ and $f$ from which we calculated the absolute
value $\delta v = \sqrt{c^2 + d^2}$ and the direction $\varphi = 90^{\circ} 
- \arctan{(d/c)}$ of the velocity gradient. In this definition the position
angle, $\varphi$, is measured in the usual (east of north) sense.
The velocity gradients are listed in 
Table~\ref{tab_properties} for those clouds where a meaningful fit could be
obtained. The method, of course, only provides us with a linear approximation
of an average global velocity gradient across each cloud, whereas the observed
gradients are not expected to be perfectly linear and, on the local scale,
will deviate from the average value and direction determined from the planar
fits.

\begin{figure}
  \includegraphics[width=\linewidth]{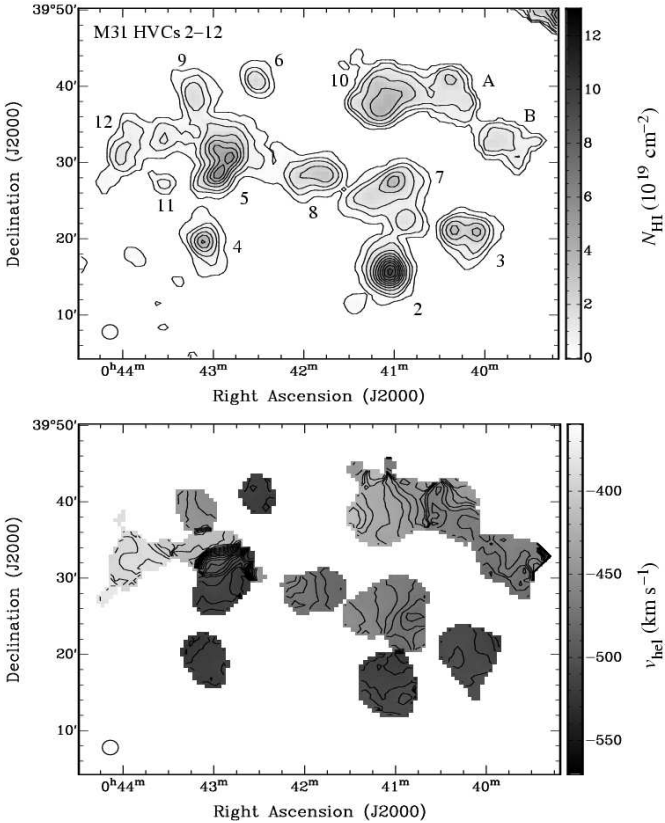}
  \caption{~Map of the \ion{H}{i} column densities (top) and the heliocentric
radial velocities (bottom) of M31~HVCs~2--12. The contours in the upper map
are drawn at $1 \cdot 10^{18} \; \mathrm{cm}^{-2}$, $5 \cdot 10^{18} \; 
\mathrm{cm}^{-2}$ and from $1 \cdot 10^{19} \; \mathrm{cm}^{-2}$ in steps of 
$1 \cdot 10^{19} \; \mathrm{cm}^{-2}$. The contours in the lower map are 
separated by $5 \; \mathrm{km \, s}^{-1}$.}
  \label{fig_ov0}
\end{figure}

The ellipticities of the HVCs were obtained by fitting an ellipse to the
column density distribution. The parameters of the ellipse were derived from a
weighted second moment analysis, following the procedure described by Banks et
al. (\cite{Banks95}). The ellipse centre was fixed by the first moment of the 
column density distribution. From the calculation of the second moments we then 
derived the major and minor axis, $a$ and $b$, of the ellipse. The ellipticities 
given in Table~\ref{tab_properties} are defined by $e = 1 - \frac{b}{a}$. 
Only \object{M31~HVC~12} was so diffuse and extended that a meaningful fit could not be
obtained.

\subsection{\object{M31 HVC 1}}

\object{M31~HVC~1} is located about half a degree south of M31~HVCs~2--12. This
proximity together with similar radial velocities suggests a connection with
M31~HVCs~2--12. \object{M31~HVC~1} reveals the highest peak column density detected
among all clouds in our sample with $N_{\mathrm{HI}} = 1.1 \cdot 10^{20} \;
\mathrm{cm}^{-2}$ at $2'$ resolution. The \ion{H}{i} mass is $M_{\mathrm{HI}}
= 5 \cdot 10^5 \, M_{\odot}$. Maps and profiles of the cloud are presented in
Fig.~\ref{fig_ov1}. At first glance, \object{M31~HVC~1} has a spherically-symmetric
appearance. Furthermore, it shows a systematic radial velocity gradient,
measuring about $15 \; \mathrm{km \, s}^{-1}$ in an approximately north-south
direction. The high brightness of \object{M31~HVC~1} allowed us to additionally
generate a map with $30''$ angular resolution which is shown in
Fig.~\ref{fig_hr1}. The high-resolution column density map resolves \object{M31~HVC~1}
into a fairly compact core with a peak column density of more than $2 \cdot
10^{20} \; \mathrm{cm}^{-2}$ which is embedded in a more diffuse envelope with
a conspicuous extension in the northern direction. We also produced a data
cube with higher velocity resolution which shows that the spectral lines in
the direction of the core have significantly smaller line widths of only about
$6 \; \mathrm{km \, s}^{-1}$, indicating the presence of cold gas in the core.

\subsection{\label{sec_m31_2}M31 HVCs 2--12}

\begin{figure*}
  \begin{center}
    \includegraphics[width=\linewidth]{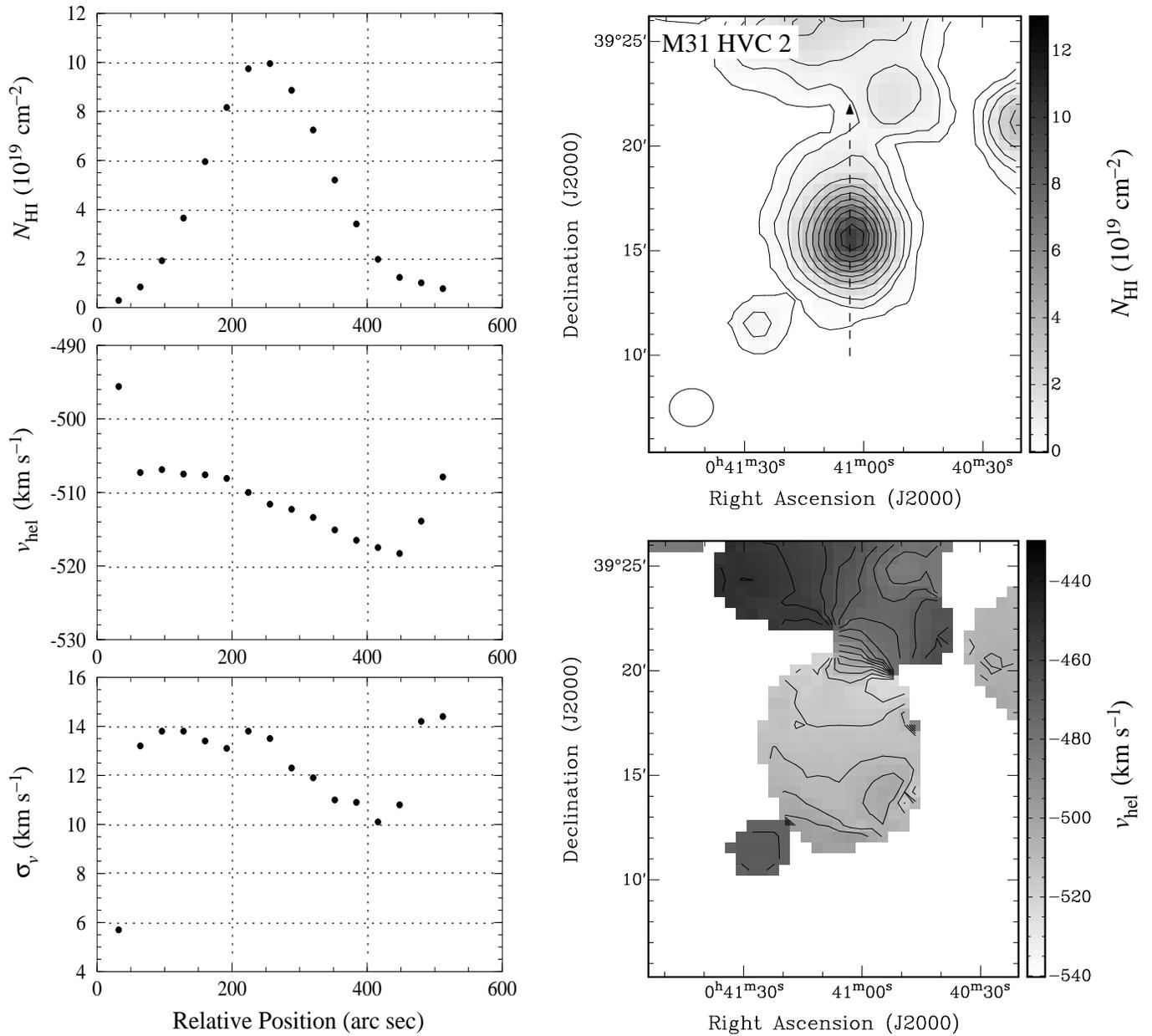}
  \end{center}
\caption{\object{M31~HVC~2}. The upper right map displays the \ion{H}{i} column
density. The contours are drawn at $1 \cdot 10^{18} \; \mathrm{cm}^{-2}$, 
$5 \cdot 10^{18} \; \mathrm{cm}^{-2}$, and
from $1 \cdot 10^{19} \; \mathrm{cm}^{-2}$ in steps of $1 \cdot 10^{19} \;
\mathrm{cm}^{-2}$. The lower right map shows the heliocentric radial
velocity. The contours are separated by $5 \; \mathrm{km \, s}^{-1}$. The
diagrams on the left hand side show from top to bottom: the \ion{H}{i} column
density, the heliocentric radial velocity, and the velocity dispersion of the
gas along the cut marked by the arrow in the column density map.}
  \label{fig_ov2}
\end{figure*}

\begin{figure*}
  \begin{center}
    \includegraphics[width=\linewidth]{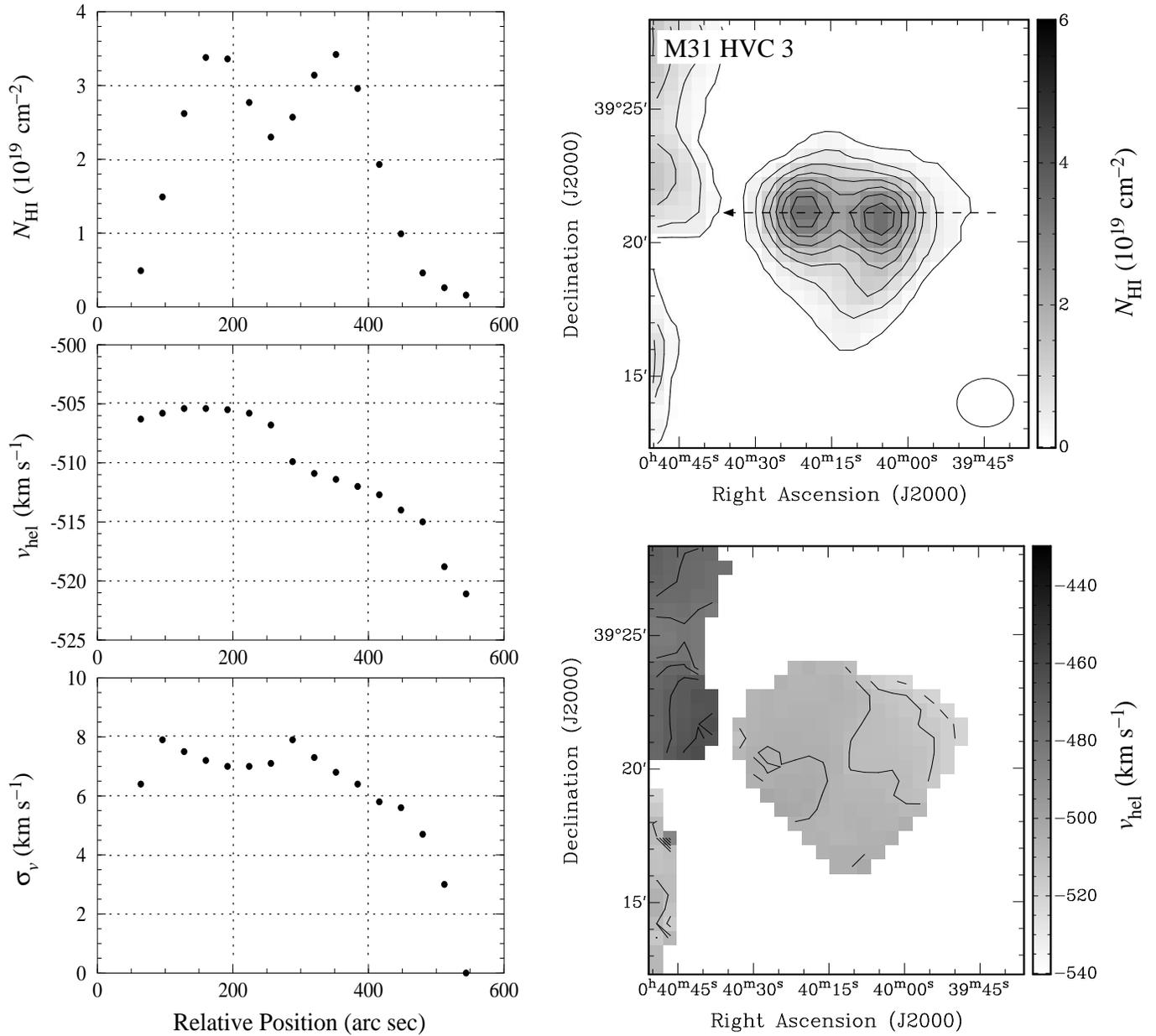}
  \end{center}
  \caption{\object{M31~HVC~3}. The upper right map displays the \ion{H}{i} column
  density. The contours are drawn at $1 \cdot 10^{18} \; \mathrm{cm}^{-2}$ and
  from $5 \cdot 10^{18} \; \mathrm{cm}^{-2}$ in steps of $5 \cdot 10^{18} \;
  \mathrm{cm}^{-2}$. The lower right map shows the heliocentric radial
  velocity. The contours are separated by $5 \; \mathrm{km \, s}^{-1}$. The
  diagrams on the left hand side show from top to bottom: the \ion{H}{i}
  column density, the heliocentric radial velocity, and the velocity
  dispersion of the gas along the cut marked by the arrow in the column
  density map.}
  \label{fig_ov3}
\end{figure*}

\begin{figure*}
  \begin{center}
    \includegraphics[width=\linewidth]{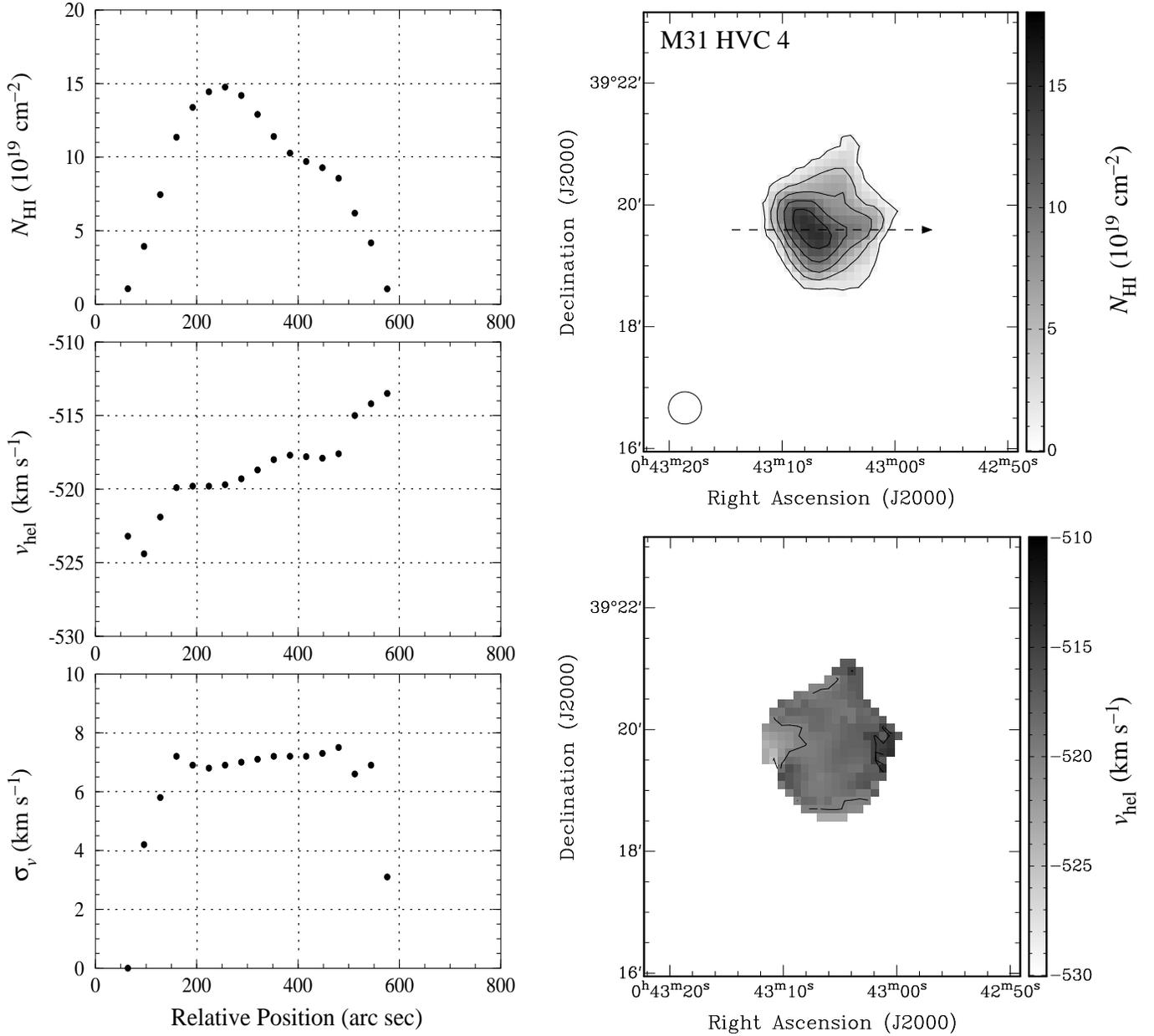}
  \end{center}
  \caption{\object{M31~HVC~4}. The upper right map displays the \ion{H}{i} column
  density. The contours are drawn from $1 \cdot 10^{19} \; \mathrm{cm}^{-2}$
  in steps of $3 \cdot 10^{19} \; \mathrm{cm}^{-2}$. The lower right map shows
  the heliocentric radial velocity. The contours are separated by $5 \;
  \mathrm{km \, s}^{-1}$. The diagrams on the left hand side show from top to
  bottom: the \ion{H}{i} column density, the heliocentric radial velocity, and
  the velocity dispersion of the gas along the cut marked by the arrow in the
  column density map.}
  \label{fig_ov4}
\end{figure*}

\begin{figure*}
  \begin{center}
    \includegraphics[width=\linewidth]{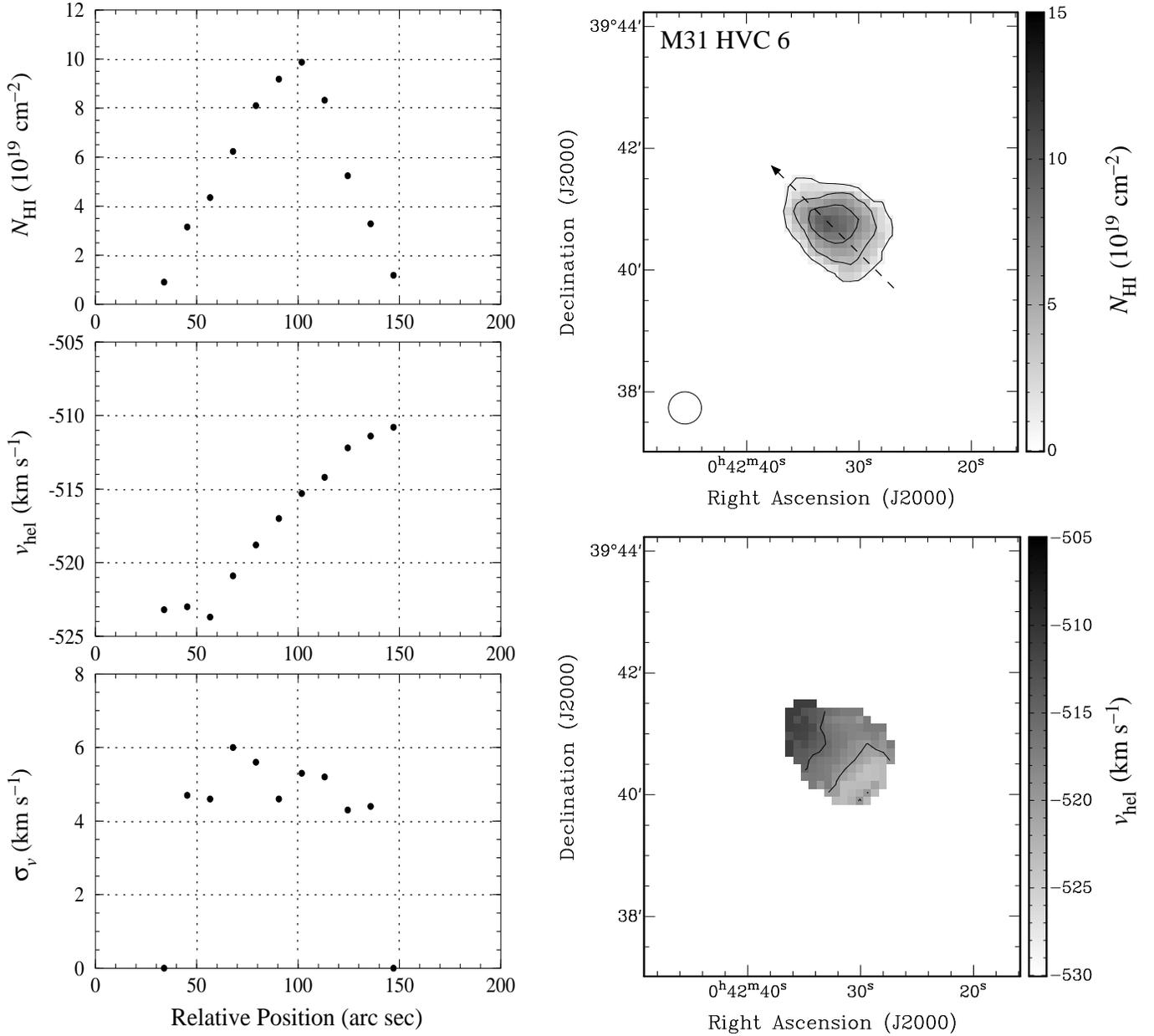}
  \end{center}
  \caption{\object{M31~HVC~6}. The upper right map displays the \ion{H}{i} column
  density. The contours are drawn from $1 \cdot 10^{19} \; \mathrm{cm}^{-2}$
  in steps of $3 \cdot 10^{19} \; \mathrm{cm}^{-2}$. The lower right map shows
  the heliocentric radial velocity. The contours are separated by $5 \;
  \mathrm{km \, s}^{-1}$. The diagrams on the left hand side show from top to
  bottom: the \ion{H}{i} column density, the heliocentric radial velocity, and
  the velocity dispersion of the gas along the cut marked by the arrow in the
  column density map.}
  \label{fig_ov6}
\end{figure*}

\begin{figure*}
  \begin{center}
    \includegraphics[width=\linewidth]{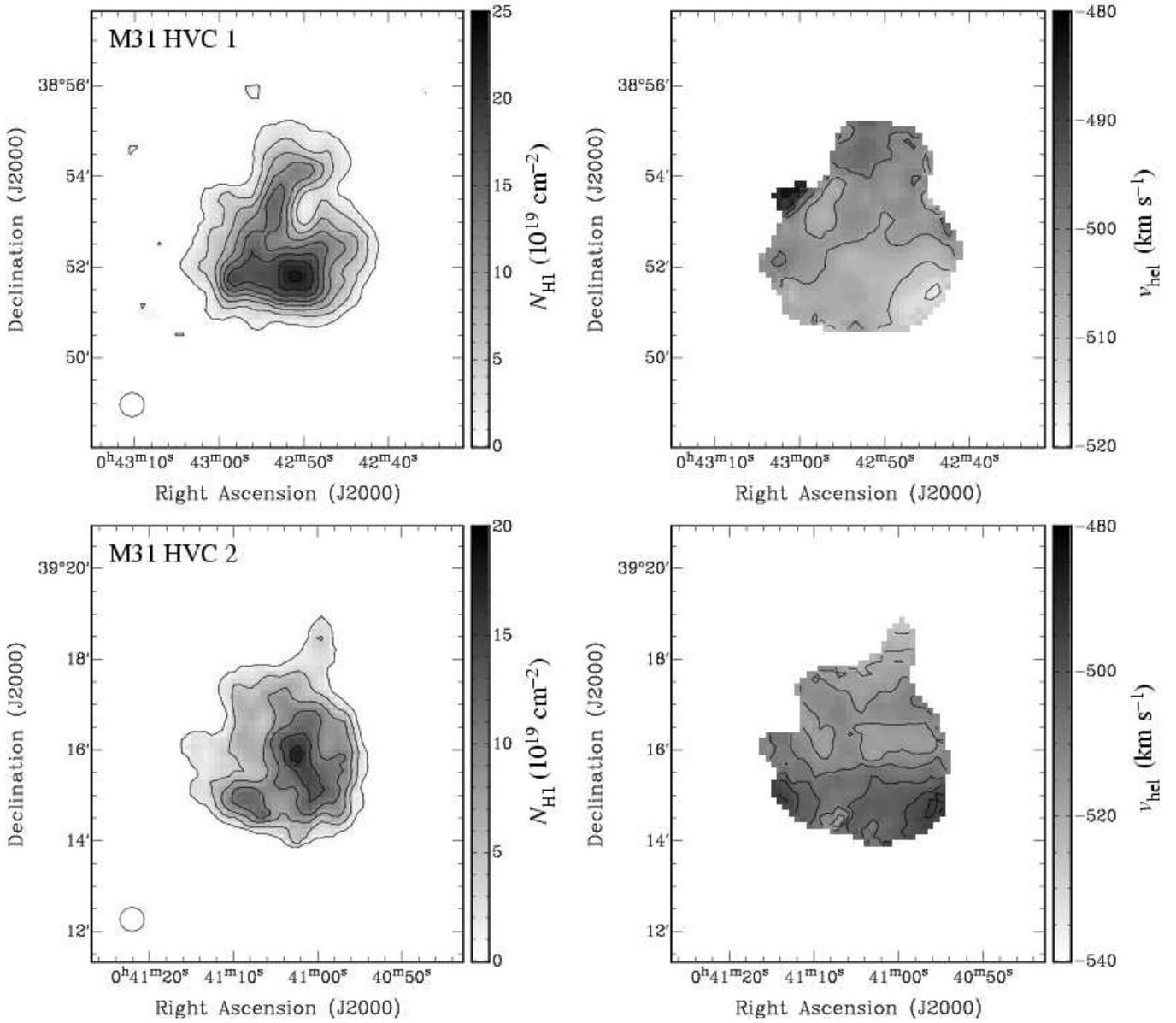}
  \end{center}
  \caption{Maps of the \ion{H}{i} column density (left) and the heliocentric
  radial velocity (right) of M31~HVCs~1 and 2 with a higher angular resolution
  of about $30''$. The contours in the column density maps are drawn from 
  $1 \cdot 10^{19} \; \mathrm{cm}^{-2}$ in steps of $3 \cdot 10^{19} \; 
  \mathrm{cm}^{-2}$. The contours in the radial velocity maps are drawn in 
  steps of $5 \; \mathrm{km \, s}^{-1}$. Both clouds show compact sub-structure 
  which is not resolved in the $2'$ maps. \label{fig_hr1}}
\end{figure*}

\begin{figure*}
  \begin{center}
    \includegraphics[width=\linewidth]{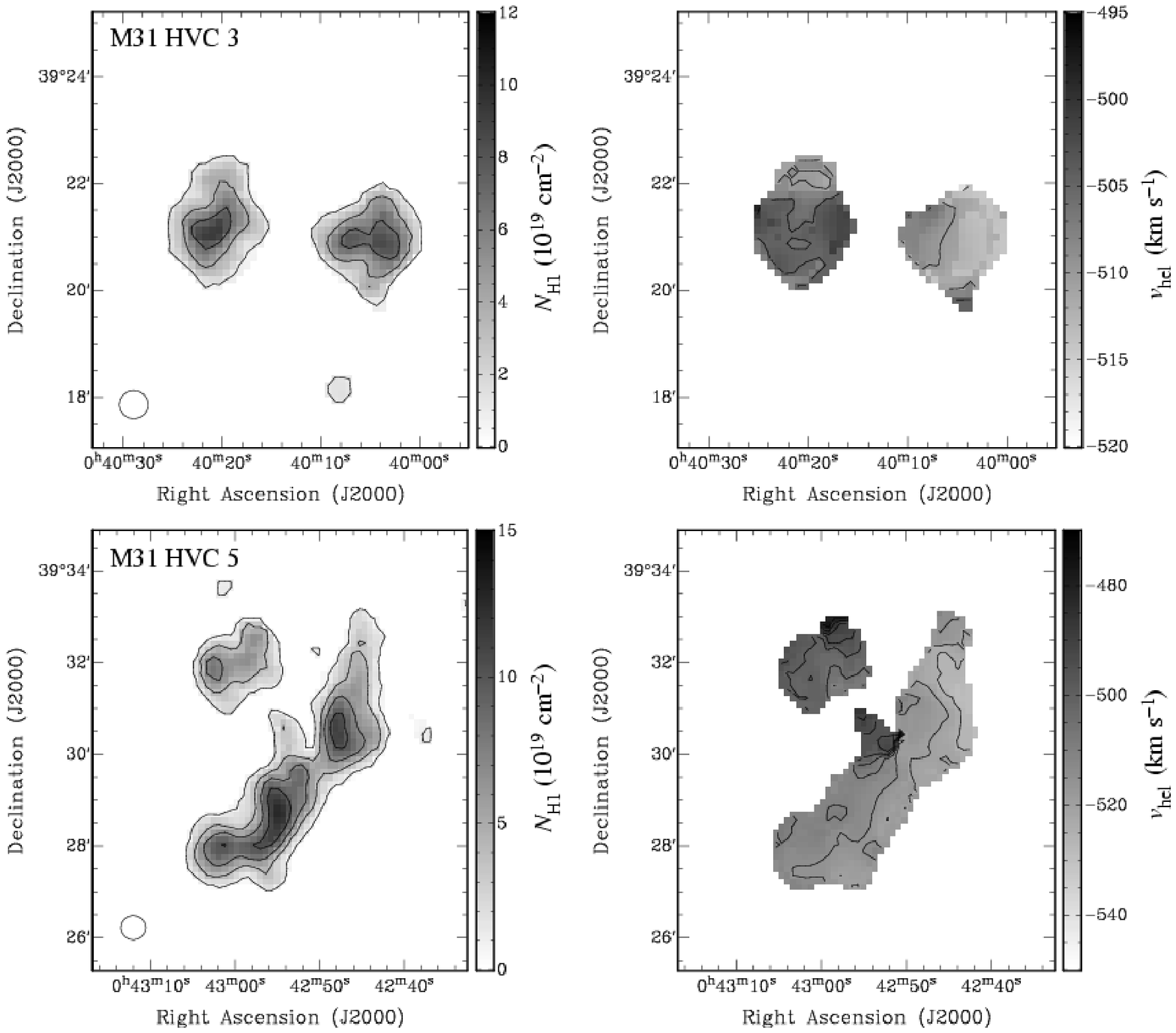}
  \end{center}
  \caption{Maps of the \ion{H}{i} column density (left) and the heliocentric
  radial velocity (right) of M31~HVCs~3 and 5 with a higher angular resolution
  of about $30''$. The contours in the column density maps are drawn from 
  $1 \cdot 10^{19} \; \mathrm{cm}^{-2}$ in steps of $3 \cdot 10^{19} \; 
  \mathrm{cm}^{-2}$. The contours in the radial velocity maps are drawn in 
  steps of $5 \; \mathrm{km \, s}^{-1}$. Both clouds break up into sub-clumps 
  which are hardly resolved in the $2'$ maps. \label{fig_hr2}}
\end{figure*}

M31~HVCs~2--12 are probably the most remarkable feature detected in our
survey. They are crowded in an area in the sky of about $1^{\circ}$ in size,
and they are located in projection close to the edge of the \ion{H}{i} disk of
\object{M31}. An overview of the whole field is shown in Fig.~\ref{fig_ov0} while maps
and profiles of some individual clouds are presented in Fig.~\ref{fig_ov2},
\ref{fig_ov3}, \ref{fig_ov4}, and \ref{fig_ov6}. Maps with $30''$ resolution
of some of the HVCs are shown in Fig.~\ref{fig_hr1} and \ref{fig_hr2}. The
individual clouds in this complex of HVCs appear to be arranged in filaments
intersecting each other at different angles. The radial velocities of the
clouds cover a very large range of $-520 \; \mathrm{km \, s}^{-1} \lesssim
v_{\mathrm{hel}} \lesssim -390 \; \mathrm{km \, s}^{-1}$, and it looks as if
clouds along the apparent filaments have similar radial velocities.

There also appears to be a morphological distinction between some of the
filaments. M31~HVCs~2--6 (and also HVC 1 which is presumably associated with
the HVC complex) have very similar radial velocities of $v_{\mathrm{hel}}
\approx -520 \ldots -500 \; \mathrm{km \, s}^{-1}$. Furthermore, they all show
high peak fluxes and mean volume densities as well as being fairly
condensed. M31~HVCs~7--12, in contrast, have different radial velocities of
$v_{\mathrm{hel}} \approx -480 \ldots {-390} \; \mathrm{km \, s}^{-1}$, and they
are more diffuse with typically lower peak fluxes and densities. The average
\ion{H}{i} volume densities and the heliocentric radial velocities of the 12
HVCs are plotted in Fig.~\ref{fig_complex}~(a) and (b). A distinction between
the two different groups of HVCs in this area can also be made on the basis of
the velocity gradients observed across the clouds. While the velocity
gradients across M31~HVCs~1--6 have arbitrary orientations, the somewhat
larger gradients observed across M31~HVCs~7--12 appear to be oriented in the
same direction (see Fig.~\ref{fig_complex}~(c)).

Two other clouds at the north-western edge of the map have been named
M31~HVC~A and B. Both are fairly diffuse with only moderate \ion{H}{i} peak
column densities and relatively broad spectral lines which distinguishes them
from the HVCs discussed before. The radial velocities along the two clouds
seem to follow those observed for the \ion{H}{i} disk of \object{M31} which was also
covered at the edge of our field of view and which can partly be seen in the
upper-right corner of the column density map in Fig.~\ref{fig_ov0}. This
suggests that M31~HVC~A and B might represent gas of an outer disk component
of \object{M31}.

\subsection{\object{M31 HVC 13}}

\begin{figure*}
  \begin{center}
    \includegraphics[width=\linewidth]{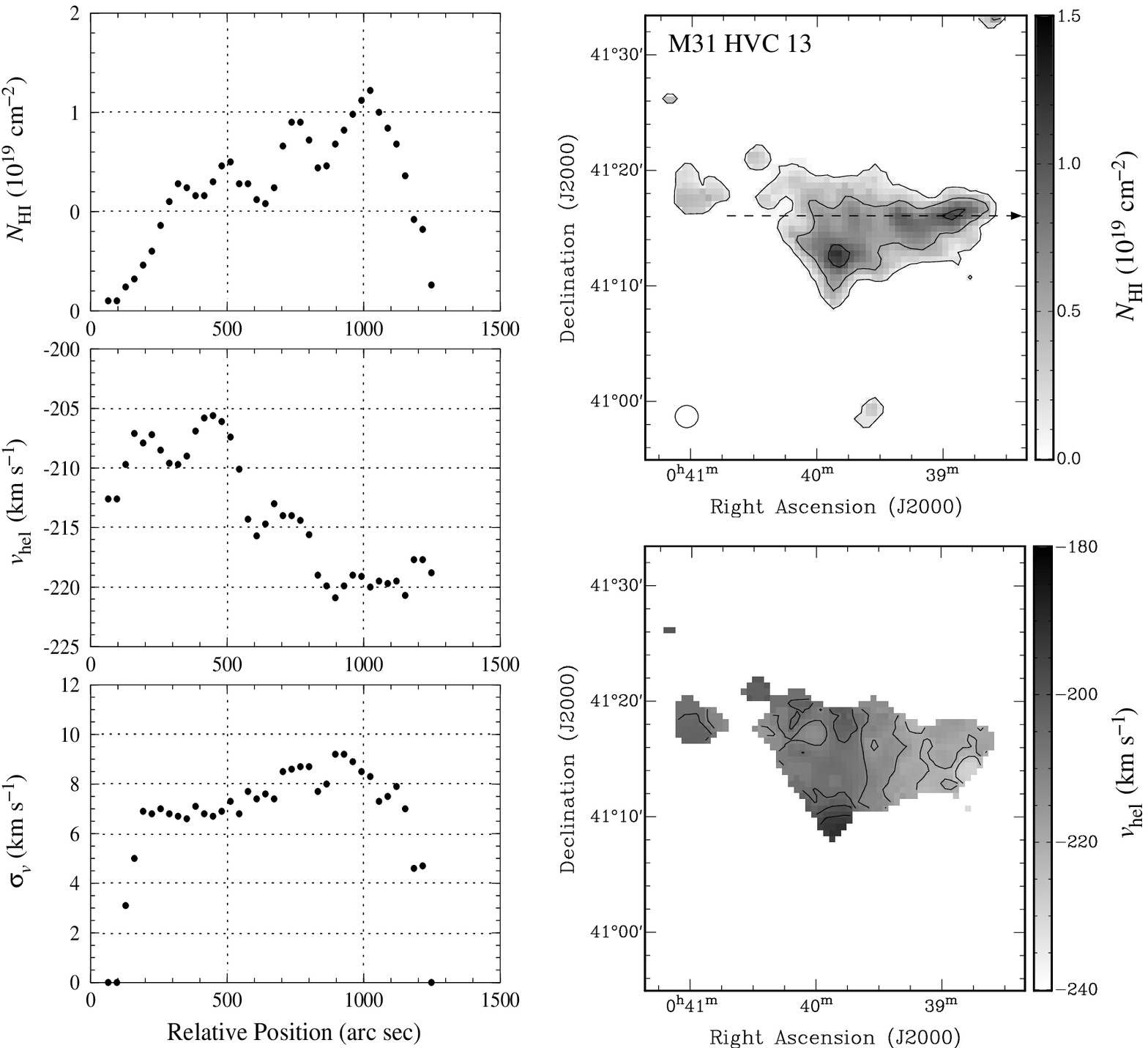}
  \end{center}
  \caption{\object{M31~HVC~13}. The upper right map displays the \ion{H}{i} column
  density. The contours are drawn at $1 \cdot 10^{18} \; \mathrm{cm}^{-2}$ and
  from $5 \cdot 10^{18} \; \mathrm{cm}^{-2}$ in steps of $5 \cdot 10^{18} \;
  \mathrm{cm}^{-2}$. The lower right map shows the heliocentric radial
  velocity. The contours are separated by $5 \; \mathrm{km \, s}^{-1}$. The
  diagrams on the left hand side show from top to bottom: the \ion{H}{i}
  column density, the heliocentric radial velocity, and the velocity
  dispersion of the gas along the cut marked by the arrow in the column
  density map.}
  \label{fig_ov13}
\end{figure*}

\object{M31~HVC~13} is rather diffuse and extended. It has quite low column densities
and appears to be fragmented into smaller clumps towards the eastern
edge. Maps and profiles are shown in Fig.~\ref{fig_ov13}. \object{M31~HVC~13} appears
in projection against the \ion{H}{i} disk of \object{M31}. But its radial velocity is
different by about $130 \; \mathrm{km \, s}^{-1}$ from that of the disk gas
observed in this direction, so that it is presumably located outside the disk
plane of \object{M31}. It lies only about half a degree south of \object{NGC~205}, the satellite
galaxy of \object{M31}. \ion{H}{i} emission from \object{NGC~205} was also detected at the edge
of our field of view. Interestingly, the radial velocities of the gas observed
in \object{NGC~205} and \object{M31~HVC~13} cover essentially the same range. 
Despite the positional and velocity correspondence we do not detect
a continuous \ion{H}{i} bridge between these two objects at our sensitivity.

\subsection{M31 HVCs 14--16}

\begin{figure*}
  \begin{center}
    \includegraphics[width=\linewidth]{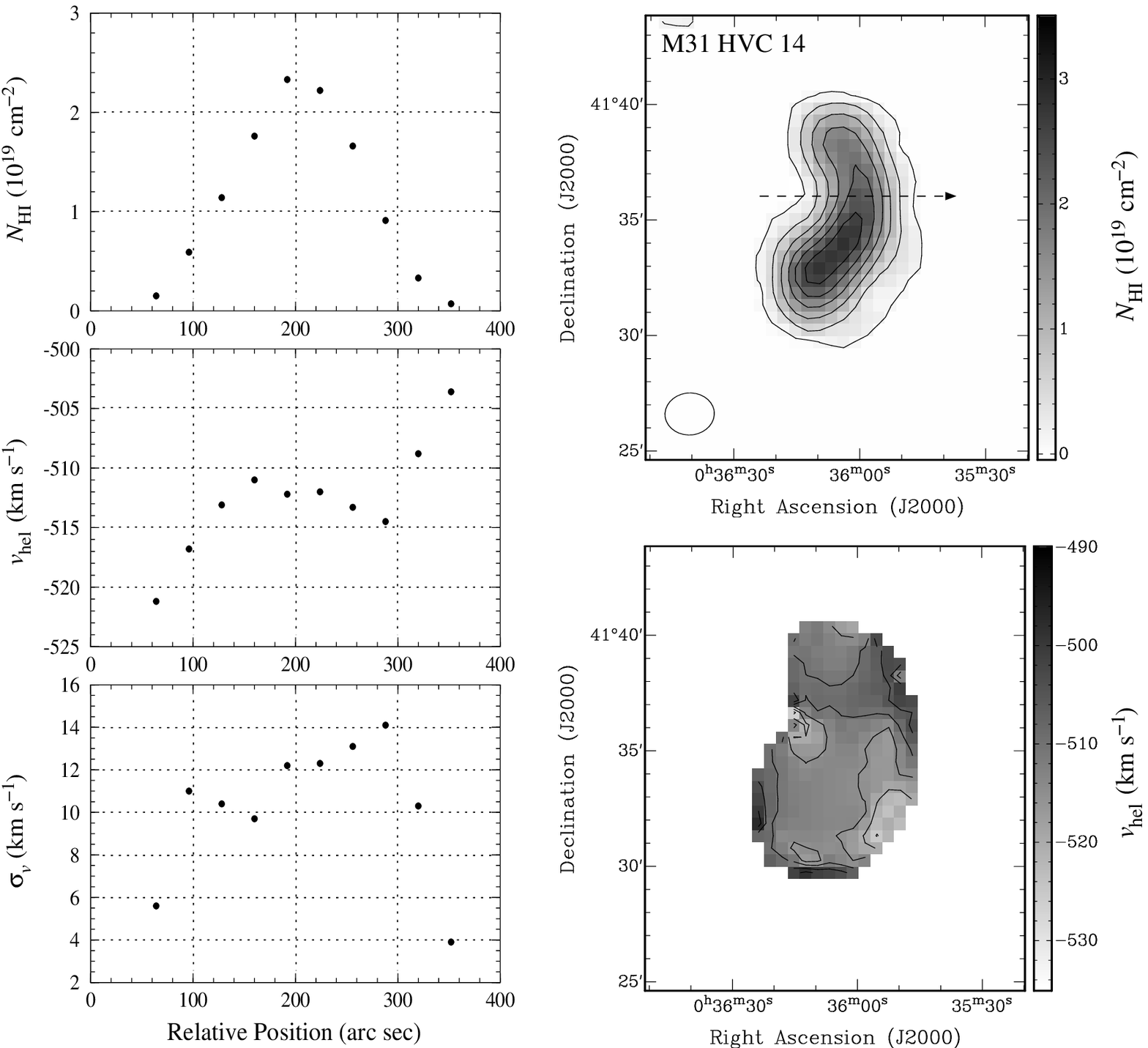}
  \end{center}
  \caption{\object{M31~HVC~14}. The upper right map displays the \ion{H}{i} column
  density. The contours are drawn at $1 \cdot 10^{18} \; \mathrm{cm}^{-2}$ and
  from $5 \cdot 10^{18} \; \mathrm{cm}^{-2}$ in steps of $5 \cdot 10^{18} \;
  \mathrm{cm}^{-2}$. The lower right map shows the heliocentric radial
  velocity. The contours are separated by $5 \; \mathrm{km \, s}^{-1}$. The
  diagrams on the left hand side show from top to bottom: the \ion{H}{i}
  column density, the heliocentric radial velocity, and the velocity
  dispersion of the gas along the cut marked by the arrow in the column
  density map.}
  \label{fig_ov14}
\end{figure*}

\begin{figure*}
  \begin{center}
    \includegraphics[width=\linewidth]{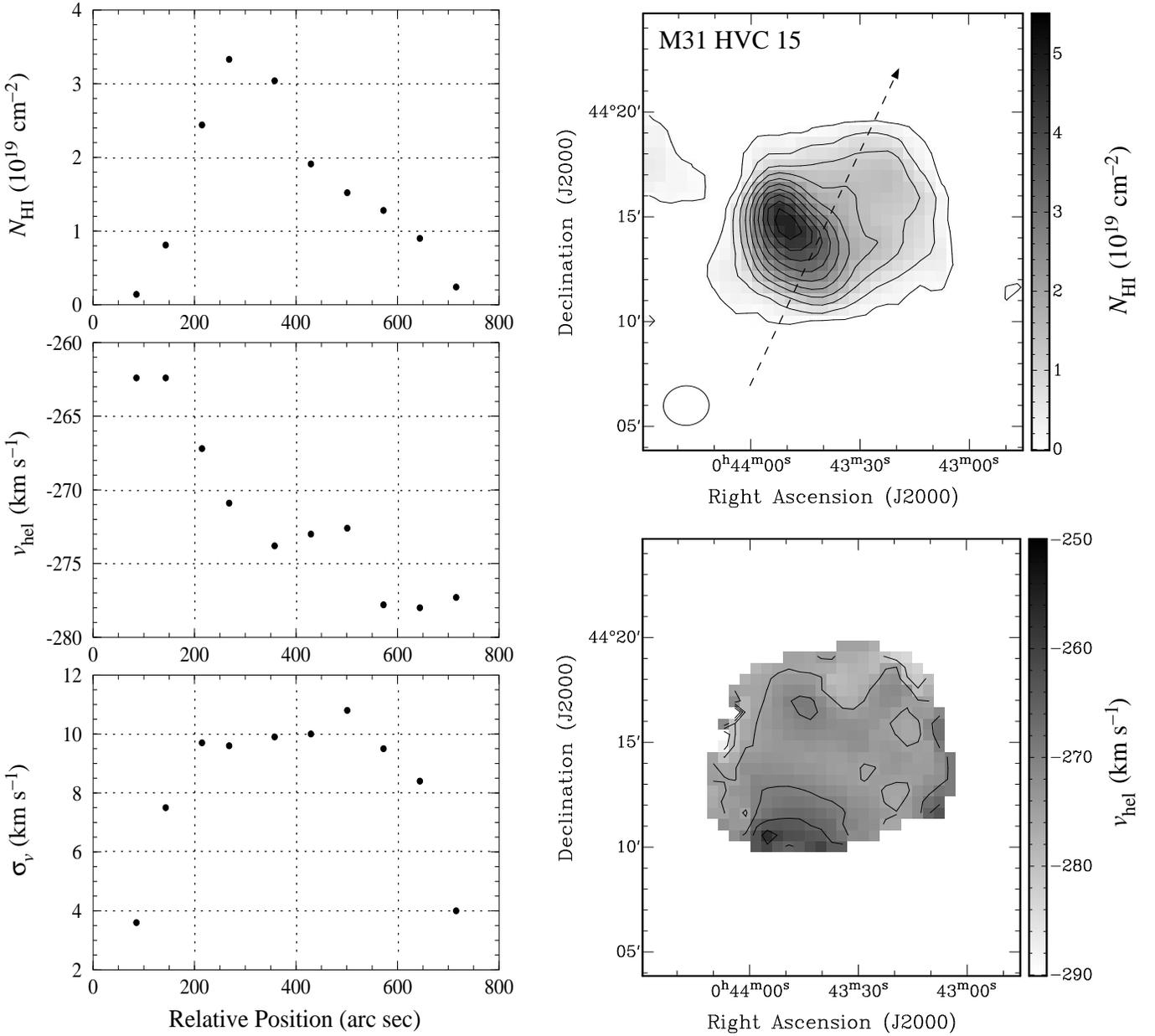}
  \end{center}
  \caption{\object{M31~HVC~15}. The upper right map displays the \ion{H}{i} column
  density. The contours are drawn at $1 \cdot 10^{18} \; \mathrm{cm}^{-2}$ and
  from $5 \cdot 10^{18} \; \mathrm{cm}^{-2}$ in steps of $5 \cdot 10^{18} \;
  \mathrm{cm}^{-2}$. The lower right map shows the heliocentric radial
  velocity. The contours are separated by $5 \; \mathrm{km \, s}^{-1}$. The
  diagrams on the left hand side show from top to bottom: the \ion{H}{i}
  column density, the heliocentric radial velocity, and the velocity
  dispersion of the gas along the cut marked by the arrow in the column
  density map.}
  \label{fig_ov15}
\end{figure*}

\begin{figure*}
  \begin{center}
    \includegraphics[width=\linewidth]{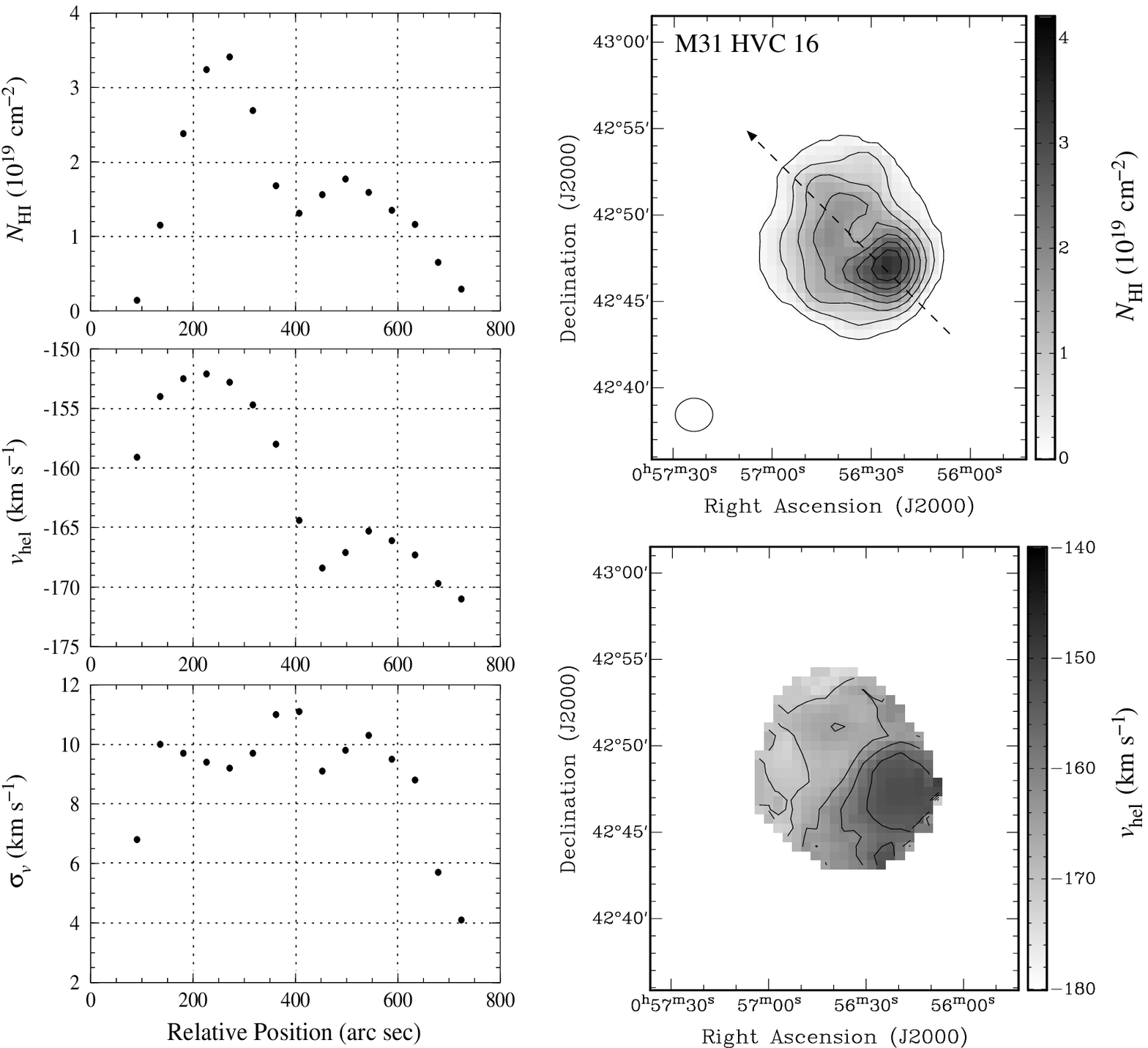}
  \end{center}
  \caption{\object{M31~HVC~16}. The upper right map displays the \ion{H}{i} column
  density. The contours are drawn at $1 \cdot 10^{18} \; \mathrm{cm}^{-2}$ and
  from $5 \cdot 10^{18} \; \mathrm{cm}^{-2}$ in steps of $5 \cdot 10^{18} \;
  \mathrm{cm}^{-2}$. The lower right map shows the heliocentric radial
  velocity. The contours are separated by $5 \; \mathrm{km \, s}^{-1}$. The
  diagrams on the left hand side show from top to bottom: the \ion{H}{i}
  column density, the heliocentric radial velocity, and the velocity
  dispersion of the gas along the cut marked by the arrow in the column
  density map.}
  \label{fig_ov16}
\end{figure*}

The remaining three HVCs observed near \object{M31} are all positionally isolated from
each other and from the disk of \object{M31}. Maps and profiles are shown in
Fig.~\ref{fig_ov14}, \ref{fig_ov15}, and \ref{fig_ov16}. All three clouds
clearly deviate from a spherically-symmetric appearance. While \object{M31~HVC~14} is
elongated and slightly bent, \object{M31~HVCs~15} and 16 exhibit a pronounced head-tail
structure. A compact core appears to be embedded in an asymmetric envelope
which forms an extended tail in one direction. Head-tail structures like these
are also widely observed in the case of Galactic HVCs (Br\"uns et
al. \cite{Bruens00}; Br\"uns et al. \cite{Bruens01}; Westmeier et
al. \cite{Westmeier05b}), and they are suggestive of a distortion of the
clouds by external forces like e.g. the ram pressure of an ambient medium.

\subsection{\object{M33 HVC 1}}

\begin{figure*}
  \begin{center}
    \includegraphics[width=\linewidth]{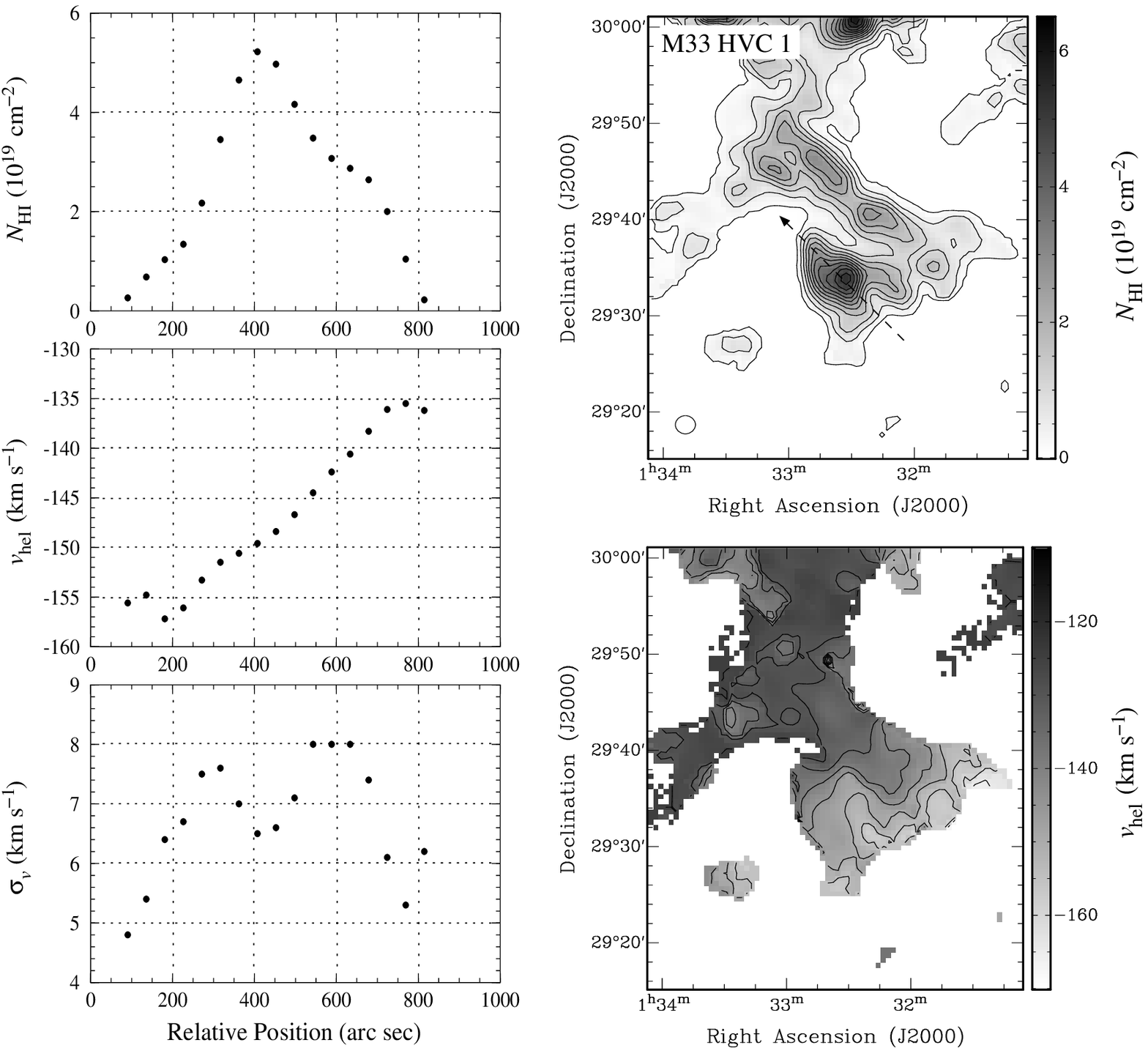}
  \end{center}
  \caption{\object{M33~HVC~1}. The upper right map displays the \ion{H}{i} column
  density. The contours are drawn at $1 \cdot 10^{18} \; \mathrm{cm}^{-2}$ and
  from $5 \cdot 10^{18} \; \mathrm{cm}^{-2}$ in steps of $5 \cdot 10^{18} \;
  \mathrm{cm}^{-2}$. The lower right map shows the heliocentric radial
  velocity. The contours are separated by $5 \; \mathrm{km \, s}^{-1}$. The
  diagrams on the left hand side show from top to bottom: the \ion{H}{i}
  column density, the heliocentric radial velocity, and the velocity
  dispersion of the gas along the cut marked by the arrow in the column
  density map.}
  \label{fig_ovA}
\end{figure*}

\object{M33~HVC~1} is located about $1^{\circ}$ south of the centre of \object{M33}. It is the
most conspicuous HVC around \object{M33} found in the GBT survey by Thilker et al. (in 
prep.). In the GBT data it is already obvious that the HVC is not isolated but 
connected with the \ion{H}{i} disk of \object{M33} by a faint gas bridge. In our WSRT 
data (Fig.~\ref{fig_ovA}) this bridge is very prominent with \ion{H}{i} column
densities of a few times $10^{19} \; \mathrm{cm}^{-2}$. Both \object{M33~HVC~1} and the
gas bridge share a common gradient in radial velocity measuring about $20 \;
\mathrm{km \, s}^{-1}$.

\section{\label{sect_discussion}Discussion}

\subsection{\label{sect_stability}Dynamical mass and stability}

\begin{table}
  \caption{Parameters for the determination of the dynamical masses of 
  M31~HVCs~1, 2, 6, 8, and 10. $\delta v$ is the observed radial velocity gradient 
  across the angular extent $\vartheta$ of each cloud as extracted from the 
  one-dimensional cuts in Fig.~\ref{fig_ov1}, \ref{fig_ov2}, and \ref{fig_ov6} 
  and from the map in Fig.~\ref{fig_ov0}. $M_{\mathrm{dyn}}$ denotes the lower 
  limit for the derived dynamical mass. 
  \label{tab_dynmass}}
  \begin{center}
  \begin{tabular}{lrrr}
    \hline \hline
    object & $\delta v$                & $\vartheta$ & $M_{\mathrm{dyn}}$    \\
           & ($\mathrm{km \, s}^{-1}$) & (arc sec)   & ($10^5 \; M_{\odot}$) \\
    \hline
    \object{M31 HVC 1}  & 14 & 400 & $>\! 86$ \\
    \object{M31 HVC 2}  & 10 & 400 & $>\! 44$ \\
    \object{M31 HVC 6}  & 13 & 100 & $>\! 19$ \\
    \object{M31 HVC 8}  & 35 & 540 & $>\!730$ \\
    \object{M31 HVC 10} & 37 & 540 & $>\!810$ \\
    \hline
  \end{tabular}
  \end{center}
\end{table}

Three clouds, M31~HVCs~1, 2, and 6, have a spherical appearance combined with a
pronounced gradient in radial velocity. Two more clouds, M31~HVCs~8 and 10, 
have a pronounced, linear velocity gradient, although they appear slightly 
elliptical. If we assume these velocity gradients to be caused by a rotation of 
the clouds we can estimate their dynamical mass
\begin{equation}
  M_{\rm dyn} = \frac{R \, v_{\rm rot}^2}{G}
  \label{eqn_dyn} 
\end{equation}
where $R$ is the radius of the cloud, $v_{\rm rot}$ is the rotation velocity
at the edge, and $G$ denotes the gravitational constant. The velocity gradients 
$\delta v$ over the angular extent $\vartheta$ were estimated from the 
one-dimensional cuts across the clouds and are summarised in 
Table~\ref{tab_dynmass}. Because of the unknown inclination angle $i$ of the 
clouds the observed velocity gradient $\delta v$ only gives us a lower limit 
of the rotation velocity with $v_{\rm rot} = \delta v / (2 \cos i)$. Nonetheless, 
the obtained lower limits for the dynamical masses of the five clouds are already 
by one or two orders of magnitude larger than the observed \ion{H}{i} masses (see
Table~\ref{tab_properties}). If the assumption of rotation is correct we are 
seeing only a small fraction of the clouds' masses in neutral atomic hydrogen. 
Additional mass components like ionised gas or dark matter are required to account 
for the observed velocity gradients.

A similar result can be derived by applying the virial theorem, assuming
spherical symmetry and, for simplicity, a constant mass density. We can
justify the assumption of virialisation by making a simple estimate of the
dynamical timescales of the clouds. If we consider the mean diameter of the
clouds to be $D \approx 1 \; \mathrm{kpc}$ and the mean FWHM of the spectral
lines to be $\Delta v \approx 25 \; \mathrm{km \, s}^{-1}$ we obtain an
estimate for the dynamical timescale of $\tau_{\mathrm{dyn}} = D / \Delta v
\approx 4 \cdot 10^7$~yr. This timescale is short compared with the expected
orbital timescales of the HVCs relative to \object{M31}, about $10^9$~yr, over which
changing tidal forces might be active. Thus, we can assume that the HVCs are
at least close to internal dynamical equilibrium so that we can apply the
virial theorem which, in our case, reads
\begin{equation}
  \frac{\Delta v^2}{8 \ln 2} = \frac{G M_{\mathrm{vir}}}{5 R} \, . 
  \label{eqn_vir} 
\end{equation}
Here, $M_{\mathrm{vir}}$ is the virial mass of the cloud, $R$ is the cloud
radius, $G$ denotes the gravitational constant, and $\Delta v$ is the FWHM of
the \ion{H}{i} lines. Solving Eq.~\ref{eqn_vir} for the virial mass and
inserting the parameters given in Table~\ref{tab_properties} yields virial
masses of the order of a few times $10^7 \, M_{\odot}$ for the most condensed
clouds in our sample. The calculated virial masses lead to a typical mass
fraction of neutral hydrogen of less than 1\%. Although these are small
values, we have to consider that the observed line widths are probably only
upper limits for the velocity dispersion of the gas since internal kinematics
or turbulence can result in an additional broadening of the spectral lines.

It should also be stressed that the application of the virial theorem to
\ion{H}{i} clouds is open to justified criticism. Not only does the calculation 
of virial masses depend upon a number of assumptions with respect to the object
symmetry, but the simple form of Eq.~\ref{eqn_vir} also demands a homogeneous 
population of particles with identical masses. It is doubtful that the velocity 
dispersion of the \ion{H}{i} component alone would provide a realistic estimate 
of the total mass of HVCs. Moreover, the calculation of dynamical masses 
according to Eq.~\ref{eqn_dyn} is in conflict with the virial theorem in the 
form of Eq.~\ref{eqn_vir}. The virial theorem requires random motions of the 
particles instead of a regular rotation of the cloud. Either the dynamical mass 
or the virial mass of a particular cloud can be correct but not both. By 
calculating virial masses, we can only expect to assess the order of magnitude 
of the entire mass of a cloud which, in our case, confirms the necessity of a 
large amount of additional mass such as ionised gas or dark matter to stabilise 
the HVCs observed in our survey.

\subsection{The origin of the HVCs}

One possible origin of the HVCs observed around \object{M31} and \object{M33} could be tidal
stripping of gas during close encounters with dwarf galaxies. Thilker et
al. (\cite{Thilker04}) already noticed that the complex of HVCs found south of
the disk of \object{M31} is partly overlapping with the giant stellar stream of \object{M31}
discovered by Ibata et al. (\cite{Ibata01}). A comparison between the optical
map of the stellar stream by Ferguson et al. (\cite{Ferguson04}) and our WSRT
\ion{H}{i} data (Fig.~\ref{fig_stellarstream}) shows that the HVCs are grouped
along the southern edge of the stream close to the disk of \object{M31}. This
positional overlap suggests a connection between the stellar stream and the
\ion{H}{i} gas. Ferguson et al. (\cite{Ferguson04}) also present the results
of radial velocity measurements of the stellar component at four positions
along the stream. The heliocentric radial velocities range from $-550 \;
\mathrm{km \, s}^{-1}$ close to the disk of \object{M31} up to $-325 \; \mathrm{km \,
s}^{-1}$ about $4^{\circ}$ away from the disk. In the region of the \ion{H}{i}
gas, the radial velocities of the stellar component are in the range of about
$-480 \; \mathrm{km \, s}^{-1}$ to $-550 \; \mathrm{km \, s}^{-1}$ which is
partly overlapping with the velocities found for the HVCs in our survey. But,
as noticed in Sect.~\ref{sec_m31_2}, the velocity structure of the HVC complex
is very complicated. Several filaments of clouds with different radial
velocities appear to intersect each other at various angles, making a
comparison with the velocities of the stellar stream challenging.

In addition to having a degree of organization into filamentary structures, it
is striking that many of the \ion{H}{i} clumps, particularly HVCs 2--6 seen
in Fig.~\ref{fig_ov0}, have large internal velocity gradients which are
typically not directed along the filaments. This circumstance suggests that
even within this region of likely overall tidal origin, there may well be
dynamical effects related to localised dark matter concentrations.

Another example of a cloud that is likely of tidal origin is \object{M31~HVC~13} which
is located about half a degree south of \object{NGC~205} and covers the same radial
velocities as the gas observed towards this satellite galaxy of \object{M31}. A
possible tidal interaction between \object{NGC~205} and \object{M31} was already suggested by
Zwicky (\cite{Zwicky59}), based on the discovery of extended, faint arms
emanating from both ends of \object{NGC~205}. Recently, McConnachie et
al. (\cite{McConnachie04b}) reported the discovery of an arc-like feature in
the distribution of blue red giant branch stars north of \object{NGC~205} which might
be part of a stellar stream originating from a tidal interaction with
\object{M31}. This presumable tidal trail, however, is located on the opposite side of
\object{NGC~205} with respect to \object{M31~HVC~13}. Our current sensitivity has not permitted
detection of a continuous \ion{H}{i} bridge between the two objects, so a
physical association remains circumstantial.

We have already noted that \object{M33~HVC~1} is not isolated but appears to be
connected with the \ion{H}{i} disk of \object{M33} by a faint gas bridge. This
possible connection is suggestive of a tidal origin of \object{M33~HVC~1} although the
potential progenitor remains unkown. The distribution of stellar sources in the 
INT WFC survey of \object{M33} (McConnachie et al. \cite{McConnachie04a}) appears to show 
no significant enhancement toward \object{M33~HVC~1} or its associated \ion{H}{i} bridge. 
This suggets the progenitor of \object{M33~HVC~1} either: (1) is quite low in stellar mass 
relative to the Andromeda stream, (2) lies outside the INT survey field, or (3) 
has been completely integrated by \object{M33}. It is also possible that a progenitor never 
existed, that is, \object{M33~HVC~1} never supported star formation and remains largely 
intact as a gaseous, primordial dark matter halo. Thus, the region around \object{M33~HVC~1} 
remains a promising target for deep optical surveys which could uncover 
peculiarities in the stellar distribution associated with the high-velocity 
\ion{H}{i} gas.

The remaining three HVCs studied near \object{M31} are all isolated in
position-velocity space from both the stellar stream and any known dwarf
companion of \object{M31}, making a tidal origin questionable. Another possible origin
for the HVCs arises from current CDM structure formation scenarios, predicting
a hierarchical formation of dark-matter haloes. Large galaxies like the Milky
Way or \object{M31} are believed to have formed by accreting smaller dark-matter haloes
and thereby growing to their current mass and size. Numerical simulations
predict a large number of so-called dark-matter mini-haloes throughout the
Local Group to be left over from this hierarchical process of galaxy formation
(Klypin et al. \cite{Klypin99}; Moore et al. \cite{Moore99}).

Recent numerical simulations by Kravtsov et al. (\cite{Kravtsov04}) confirmed
the significant overabundance of dwarf-like dark-matter haloes in the models
in comparison to the relatively small number of observed dwarf galaxies around
the Milky Way. They also included a simple model of star formation in
dark-matter haloes which, for the first time, can reproduce both the circular
velocity function as well as the radial distribution of dwarf galaxies around
the Milky Way. For the central 50~kpc around \object{M31} the model of Kravtsov et
al. (\cite{Kravtsov04}) predicts only 2--5 dark-matter haloes with associated
gas masses, $M_{\mathrm{g}} > 10^6 \, M_{\odot}$, that have not undergone
significant internal star formation. They concur with the conclusion of
Thilker et al. (\cite{Thilker04}) that only a subset of the HVCs found near
\object{M31} might represent primordial dark-matter haloes while the remaining clouds
might be the result of tidal interaction. Our new imaging results support
these conclusions. A recent complementary \ion{H}{i} blind survey for HVCs
around \object{M31} with the Effelsberg telescope (Westmeier et al. \cite{Westmeier05a}), 
however, yields no evidence for any further compact HVCs beyond the boundaries 
of the GBT map of Thilker et al. (\cite{Thilker04}), although extended, 
filamentary \ion{H}{i} emission was found by Braun \& Thilker (\cite{Braun04}) 
even at larger projected distances from \object{M31} with the WSRT. The Effelsberg survey 
has a $3 \sigma$ \ion{H}{i} mass limit of about $6 \cdot 10^4 \, M_{\odot}$. It 
reaches out to a projected distance of about 130~kpc from \object{M31}, but it covers only 
a limited azimuthal range. The non-detection of additional compact \ion{H}{i} 
clouds may be an indication that the thermodynamic considerations of Sternberg et 
al. (\cite{Sternberg02}) may apply. Their calculations suggest that the \ion{H}{i} 
mass associated with a particular dark mini-halo mass is a sensitive function of 
the ambient pressure of the circum-galactic environment. This would imply a 
substantial gradient of decreasing neutral gas mass with galacto-centric radius 
for a given dark halo mass.

\section{\label{sect_summary}Summary and conclusions}

We have observed nine fields around \object{M31} and one field near \object{M33} in \ion{H}{i}
with the WSRT to study some of the HVCs discovered by Thilker et
al. (\cite{Thilker04}) and Thilker et al. (in prep.) with arc-minute
resolution. In two of the nine fields around \object{M31} we could not detect the
high-velocity gas, presumably because the emission is too diffuse and
faint. In the remaining seven fields we identify about 16 individual HVCs with
angular sizes of the order of $10'$, corresponding to absolute sizes of the
order of 1~kpc if a distance of 780~kpc (Stanek \& Garnavich \cite{Stanek98})
is assumed for \object{M31}. The observed \ion{H}{i} masses are in the range of about a
few times $10^4 \, M_{\odot}$ to $6 \cdot 10^5 \, M_{\odot}$. Under the
assumption of gravitational stability only a very small fraction of the order
of 1\% of the total mass of the HVCs can be seen in neutral, atomic hydrogen,
and additional mass components like ionised gas or dark matter have to be the
dominant constituents. The application of the virial theorem is only 
of marginal suitability because it is subject to a number of restrictions which, 
in many cases, are not stringently fulfilled. The estimation of the dynamical 
masses of five of the HVCs, however, emphasises the necessity of a large 
fraction of undetected mass.

Twelve of the HVCs are crowded in an area of only about $1^{\circ} \times
1^{\circ}$ at a projected separation of less than 15~kpc from the disk of
\object{M31}. This remarkable complex of HVCs appears to consist of several filaments
of clouds intersecting each other at various angles. Clouds along the apparent
filaments have similar radial velocities. A detailed comparison suggests that
the HVC complex might consist of two distinct populations of clouds which can
be distinguished by their different radial velocities, velocity gradients, and
mean densities. Moreover, the HVC complex is partly overlapping with the giant
stellar stream of \object{M31} discovered by Ibata et al. (\cite{Ibata01}), suggesting
a connection with the stellar stream and a tidal origin for some or most of
these HVCs. The densest clumps within this region have large internal velocity
gradients which are not generally associated with the orientation of nearby
diffuse filaments, possibly suggesting significant local concentrations of
dark mass. Another HVC near \object{M31} is located only half a degree south of \object{NGC~205}
and at an identical radial velocity, although the evidence for a physical
association remains circumstantial.  The remaining HVCs around \object{M31} are compact
and isolated without any obvious connection with \object{M31} or one of its known
satellite galaxies.

The HVC near \object{M33} is not isolated but appears to be connected with the
\ion{H}{i} disk of \object{M33} by a faint gas bridge. Both the HVC and the bridge
share a common radial velocity gradient. This configuration is again
suggestive of a late stage tidal origin, but the progenitor of the \ion{H}{i}
gas remains unknown.

The modest number of isolated HVC detections seem to be in agreement with the
numerical simulations of Kravtsov et al. (\cite{Kravtsov04}) which predict
only 2--5 primordial dark-matter mini-haloes within the field of the GBT
survey by Thilker et al. (\cite{Thilker04}). However, a complementary
\ion{H}{i} survey of \object{M31} with the Effelsberg telescope by Westmeier et
al. (\cite{Westmeier05a}), reaching out to a projected distance of about
130~kpc from \object{M31}, reveals no further compact HVCs beyond the boundaries of the
GBT map down to an \ion{H}{i} mass limit of about $6 \cdot 10^4 \,
M_{\odot}$. The concentration of \ion{H}{i} detections at relatively small
galacto-centric radii may be consistent with the calculations of Sternberg et
al. (\cite{Sternberg02}) which suggest a rapidly declining neutral fraction
associated with a given dark halo mass for a decreasing ambient pressure in a
circum-galactic environment.

\begin{acknowledgements}
  T. W. acknowledges support by the Deutsche Forschungsgemeinschaft (German
  Research Foundation) through project number KE757/4--1. The Westerbork
  Synthesis Radio Telescope is operated by the ASTRON (Netherlands Foundation
  for Research in Astronomy) with support from the Netherlands Foundation for
  Scientific Research (NWO).
\end{acknowledgements}

\end{document}